\documentclass[aps,prd,twocolumn,superscriptaddress,showpacs]{revtex4-1}
\usepackage[utf8]{inputenc}
\usepackage{amsmath}
\usepackage{amsfonts}
\usepackage{amssymb}
\usepackage[final]{graphicx}
\usepackage[caption = false]{subfig}
\usepackage{epstopdf}
\usepackage{tabularx}
\usepackage{mathtools}
\usepackage[mathscr]{eucal}
\usepackage{epsfig}
\usepackage{float}
\usepackage{bookmark}
\begin{document}
\title{Fermi surface instabilities of symmetry-breaking and topological types on the surface of a three-dimensional topological insulator}
\author{Subhajit Sarkar}
\email{sbhjt72@gmail.com}
\affiliation{Institute of Physics, P.O.: Sainik School, Bhubaneswar 751005, Odisha, India.}
\thanks{present address: Department of Chemistry, Ben Gurion University of the Negev, Beer Sheba 8410501, Israel. Email: subhajit@post.bgu.ac.il}
\date{\today}
\begin{abstract}\label{abs}
The emergence of the Pomeranchuk instability (PI) in a Helical Fermi liquid (HFL) residing on the surface of a three-dimensional topological insulator (3D TI) is addressed at the mean-field level. An expression for the PI condition is derived in terms of a few microscopic parameters in each angular momentum channel corresponding to a central interaction between the helical electrons. It is found that because of the presence of strong spin-orbit coupling (SOC) the Landau parameter, $\bar{F}_{l}$ corresponding to a particular angular momentum channel $l$ depends not only on the electron-electron interaction in the same channel but also interactions in $(l+1)$ and $(l-1)$ channels. The formalism automatically excludes the $l=1$ PI in the HFL where the Galilean invariance is broken because of the presence of strong SOC.  It is also found that the competing PIs can only be avoided until the appearance of $l=2$ PI. In this case, the corresponding nematic instability can even be achieved in the $l=1$ angular momentum channel of interaction between the electrons. The range of interaction between the electrons plays a pivotal role in bringing out the PIs. This is established by analysing a few realistic profiles of the interaction. Another class of instability, involving a change in the topology of the Fermi surface without breaking the rotational symmetry, is found which competes with the PIs. Quantum phase transition originating from this instability is quite similar to the Lifshitz transition but is driven by electron-electron interaction. Possible connections of this instability with experiments are also described briefly.
\end{abstract}
\maketitle
\section{Introduction}
Topological insulators (TIs) are a new class of materials which possess insulating bulk states but conducting boundary states \cite{Shen, ORV, QZ, HK}. These materials have attracted a great deal of research interest partly driven by their applicability in technology and partly because of their fundamental non-trivial characteristics. In particular, three-dimensional topological insulators (3D TIs) are non-magnetic insulators that possess metallic surface states with gap-less spectrum as a consequence of the non-trivial topology of the bulk electronic wave functions of the material \cite{FKM, HM}. 
 \paragraph*{}
  The simplest non-trivial 3D TIs, which are also called the strong TI, exhibit linear dispersion relation with the structure of a Dirac cone, and an odd number of such cones are present on the surface of the sample which is further ensured by $\mathbb{Z}_2$ topological invariant of the bulk \cite{QZ, HK, FKM, HM}. Furthermore, such a phase emerges from the strong spin-orbit coupling (SOC), leading to spin-momentum locking on the surface of 3D TIs
  \cite{FKM, HM}.
\paragraph*{}
On the surface of a 3D TI, because of the existence of an odd number of Dirac cones, the Fermi sea is a circular disk containing the Dirac point itself \cite{QZ, HK, FKM, HM}. Surface Fermions constitute a Helical Fermi gas, and with adiabatically switching on the interaction these surface Fermions lead to a Fermi liquid phase on the surface of the 3D TI which is called a Helical Fermi liquid (HFL)  \cite{LM}. Moreover, the existence of a single non-degenerate circular Fermi surface makes the HFL an effective spin-less Fermi liquid -- a behaviour whose appearance has its root in the topology of the bulk \cite{LM}. 
 Such a Helical Fermi liquid theory is applicable to a number of experimentally realized 3D TIs, for which the Fermi surface has been found to be very nearly circular. 
These are, $\text{Bi}_2 \text{Se}_3$,  $\text{Bi}_2 \text{Te}_2 \text{Se}$, $\text{Sb}_x \text{Bi}_{2-x} \text{Se}_2 \text{Te} $, $\text{Bi}_{1.5} \text{Sb}_{0.5} \text{Te}_{1.7} \text{Se}_{1.3}$, $\text{Tl}_{1-x} \text{Bi}_{1+x} \text{Se}_{2-\delta}$,  strained $\alpha-\text{Sn}$ on InSb(001), and strained HgTe, where the observations have been made mostly using angle-resolved photoemission spectroscopy (ARPES) \cite{XQ, HX, PV, NX, NB, KY, KE, BD, CO}.
\paragraph*{}
The behaviour of the HFL is described by the projected Landau parameters $\bar{F}_l$ which are similar in spirit to the Landau parameters of a conventional SU(2)/Galilean invariant Fermi liquid (SU(2)-FL) \cite{LM}. However, these Landau parameters are obtained by projecting the Landau parameters of a spin-orbit coupled (non-Galilean invariant) FL on the Fermi surface (FS) of a particular Helicity (originating from strong SOC) for non-zero chemical potential $\mu$ (therefore non-zero doping) \cite{LM}. Low energy properties of the HFL are strongly influenced by the shape of the FS. When the (repulsive) interaction between the quasi-particles near the Fermi surface increases sufficiently the system evolves into a new phase where the symmetry of the Fermi surface is broken. The resulting shape deformation instability is renowned as Pomeranchuk Instability (PI) which occurs in a particular angular momentum channel $l$, when corresponding the Landau parameter $F_l$ becomes sufficiently negative \cite{PI, AGD, Vignale, Col, Noz}. For example, in HFL the $l=2$ PI leads to a new phase exhibiting nematic order characterized by elongation (contraction) of the Fermi surface along $k_x$ ($k_y$) direction and vice versa, and more interestingly a partial breakdown of spin momentum locking (in the sense that spin and momentum are no-longer orthogonal) \cite{LYM}. In the case of SU(2)-FL such PIs have been studied in a sufficiently general frame work by taking into account a generic central Fermion-Fermion (e-e) interaction both in two and three dimensions \cite{QHS, QS}. This single framework describes PIs not only in $l=2$ but also in other angular momentum channels, and automatically excludes $l=1$ PI in a Galilean invariant Fermi liquid. The absence of $l=1$ PI corresponding to the absence of spontaneous generation of currents has attracted recent attention, and investigations reveal that while in a Galilean invariant system the Galilean invariance alone abandons this, in a non-Galilean invariant system such as the HFL, the contributions from the high energy degrees of freedom cancel this spontaneous current generation \cite{WKC, KSW}. Furthermore, previous works have focused mostly on superconducting instabilities, time-reversal symmetry breaking instabilities, and nematic instabilities \cite{LYM, OHJ, san, roy, nand, sds, grov, pon, neup, kremp, zerf, jian, xu, kim, jiang, gha, nog, sit, bah, men, gor, roy1}. However, a thorough study of the PIs is still missing. Motivated by these facts, in this paper, I investigate the PI on the surface of a 3D TI originating from a central interaction, and findings are qualitatively quite different from those corresponding to the SU(2)-FL, the reason being the presence of strong SOC. 
\paragraph*{}
In this paper, I derive a microscopic expression for the Pomeranchuk instability corresponding to the HFL residing on the surface of a 3D TI. The nature of the corresponding phase transition is described in terms of a few pertinent parameters of the theory representing the properties (including its curvature) of the e-e interaction on the Fermi surface. Most importantly, because of the presence of strong SOC on the surface of a 3D TI, the different angular momentum channels get coupled and the Landau parameter for a particular angular momentum channel depends on interaction strengths not only in the same channel but also in channels higher and lower than that.  This finding has a far-reaching implication in the nematic instability corresponding to $l=2$ channel as explained in this paper. Furthermore, over a range of parameters the Pomeranchuk transition driven by the central interaction becomes first order, thereby signaling discontinuous transition without involving any quantum critical behaviour. However, nematic instability coming from the hypothetical (non-) central interaction (when compared with the phenomenological quadrupolar interaction) falls within the parameter regime where the phase transition is of second order in nature signaling quantum critical nature of the isotropic-nematic transition at zero temperature. This is in agreement with the earlier results \cite{LYM}. Using a few models of the  interaction, I have further shown that apart from the PIs there exists another class of instability involving a change in the topology of the Fermi surface (FS) which is driven by electron-electron interaction and competes with the PIs. Moreover, the range of the interaction turns out to be a decisive factor in the appearance of both PI and topological-FS instability. In this regard, it is worthwhile to mention that the appearance of the Lifshitz transition involving a change in the topology of the Fermi surface has been reported recently in the doped Topological Crystalline Insulators (TCIs) $Pb_{1-x}Sn_{x} Se$ and $Pb_{1-x}Sn_{x} Te$ where, by changing the doping an interaction-induced phase transition involving a change in the FS topology has been achieved \cite{Ple, Gye}.  
\paragraph*{}
The paper is organized as follows. I have described the model in section \ref{mod}. In section \ref{mft}, I have described the mean-field theory and derived the microscopic expression of the projected Landau parameters. Section \ref{defo} describes the microscopic version of the PI condition, and in section \ref{pt}, I explain the nature of the Pomeranchuk phase transition and the corresponding order parameters. Section \ref{okh} describes the conditions under which the $l=2$ PI can describe the nematic instability originating quadrapolar electron-electron interaction. In section \ref{tpt}, I have described the importance of range in bringing out the PI and topological instability. Section \ref{conc} concludes the paper and discuss the implications of the results. Relevant calculations are presented in Appendices.
\section{The Model}\label{mod}
The non-interacting Hamiltonian corresponding to the gap-less states with a single Dirac cone on the surface of a three dimensional topological insulator (3D TI) is given by (in the units of $\hbar = k_B = 1$),
\begin{equation} \label{h0}
H_{0} =\int \frac{d^2 k}{(2\pi)^2} \sum_{\sigma , \sigma'} c^{\dagger} _{\mathbf{k}, \sigma} [v^{0}_{F} \hat{z}\cdot(\mathbf{\tau}_{\sigma , \sigma^{\prime}} \times \mathbf{k})-\mu]c _{\mathbf{k}, \sigma^{\prime}},
\end{equation}
where $v^{0}_{F}$ is the Fermi velocity the bare particles, $\mathbf{\tau}_{\sigma , \sigma^{\prime}}$ is the vector Pauli matrix, $(\sigma, \sigma')$ are the spin indices of the electrons, and $\mu$ is the chemical potential. The helical Fermi liquid theory corresponding to these surface states in the presence of electron-electron interaction is strictly valid when $\mu > 0$, i.e., in the doped limit \cite{LM}. One can diagonalize the above Hamiltonian by an unitary transformation,
\begin{equation}\label{psi}
\psi _{\mathbf{k} , s} =\frac{1}{\sqrt{2}} (i e^{-i \theta_{\mathbf{k}}} c _{\mathbf{k}, \uparrow} + s \; c _{\mathbf{k}, \downarrow})
\end{equation}
which transform the complex Dirac fermions $c _{\mathbf{k}, \sigma}$ to helical Dirac Fermions $\psi _{\mathbf{k} , s}$ with helicity $s=\pm$, where $\theta_{\mathbf{k}} = \tan^{-1} \frac{k_y}{k_x}$ \cite{LM}. In the above equation the helicity ``+" corresponds to all the particles with positive Fermi energy $\epsilon_{F}$ above the Dirac point and the helicity ``-" corresponds to all the particles with negative Fermi energy below the Dirac point. With chemical potential $\mu >0$ only the states above the Dirac point are available. Therefore, the inverse of the (\ref{psi}) is given by, $c _{\mathbf{k}, \sigma} = \eta_{\mathbf{k}} \psi_{\mathbf{k}, +}$ where $\eta^{\dagger}_{\mathbf{k}} = \frac{1}{\sqrt{2}} \begin{pmatrix}
-i e^{i \theta_{\mathbf{k}}} & 1
\end{pmatrix} $. 
Here the negative helicity Fermion operators are dropped altogether ,i.e., all the complex Dirac Fermions available in the system are projected onto the positive helicity Hilbert space. For ease of notation I shall denote the $\psi_{\mathbf{k}, +}$ as $\psi_{\mathbf{k}}$ without explicitly using the ``+" notation.
\begin{figure}[h]\label{disp}
\includegraphics[scale=0.4]{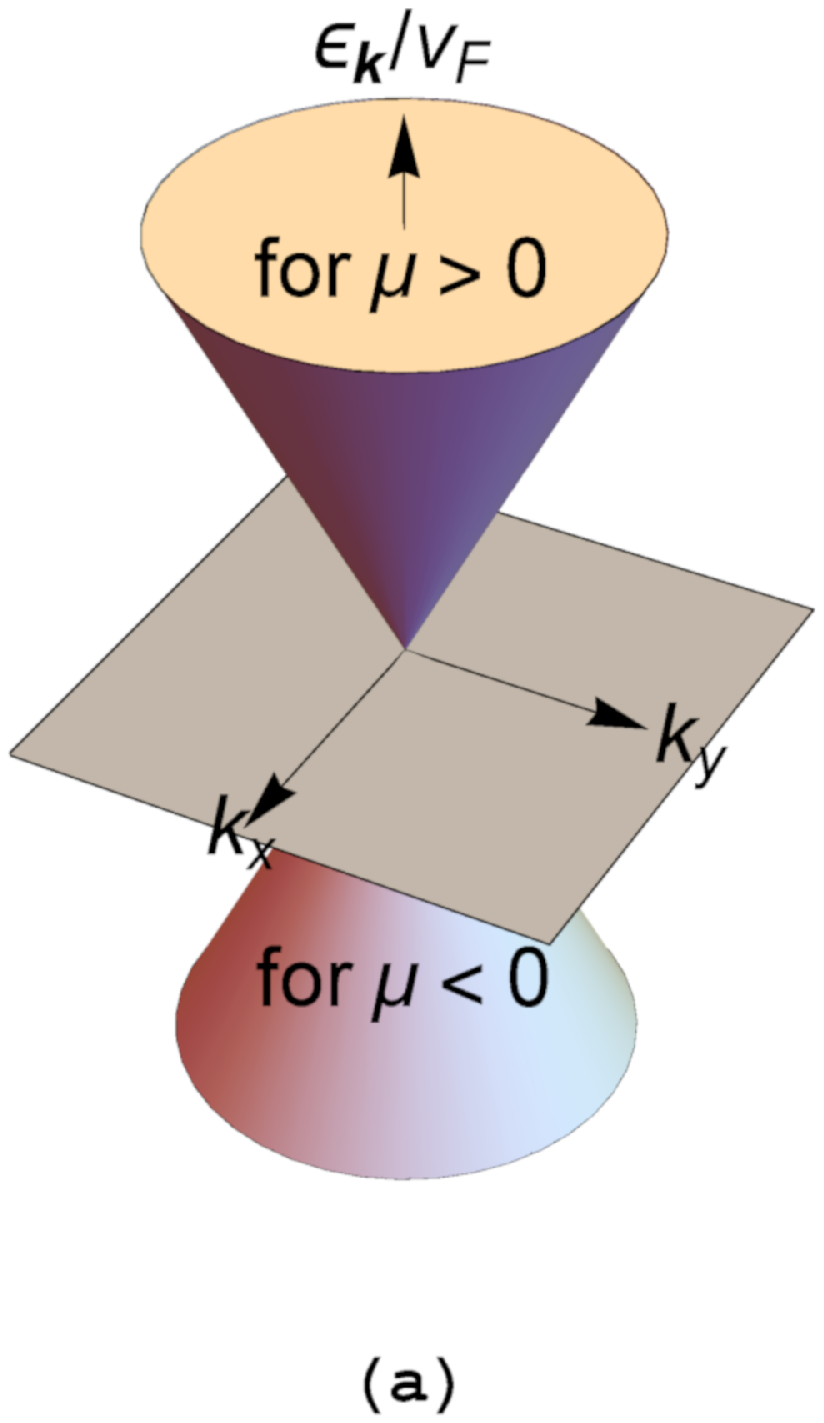}
\includegraphics[scale=0.4]{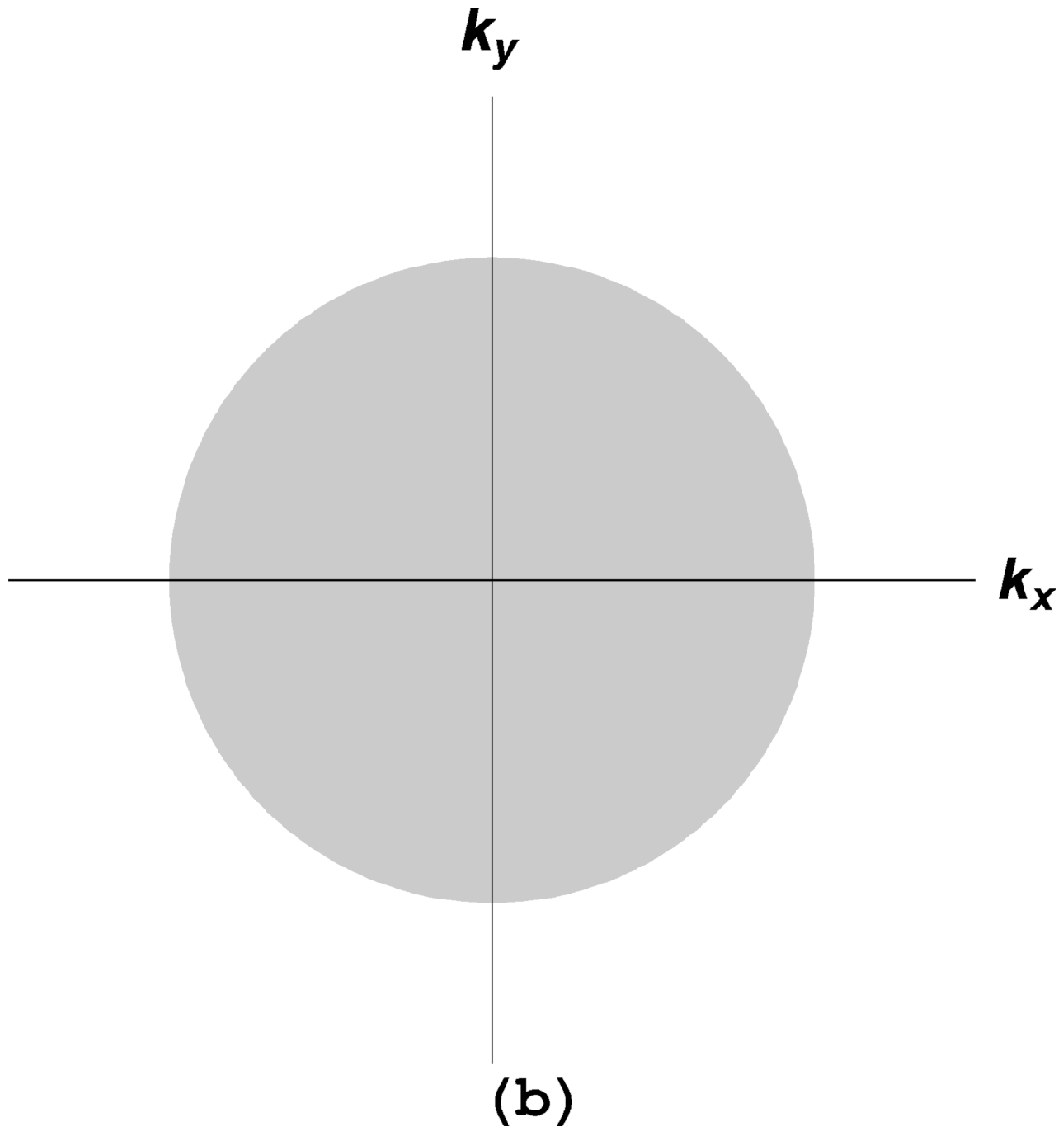}
\caption{(a) Dispersion relation $\epsilon_{\mathbf{k}}$. For $\mu > 0$ the only available states are corresponding to the Dirac cone for which  $\epsilon_{\mathbf{k}} \geq 0$. (b) The shaded circular region on $(k_x, k_y)$-plane represents the filled Fermi sea with radius $k_F = \frac{\mu}{v_{F}}$, undeformed Fermi surface being a circle.}
\end{figure}
\paragraph*{}
 The diagonalized non-interacting Hamiltonian in the projected space is then given by,
\begin{equation}\label{h01}
H_{0} = \int \frac{d^2 k}{(2\pi)^2} (\epsilon_{\mathbf{k}} - \mu) \psi^{\dagger}_{\mathbf{k}} \psi_{\mathbf{k}},
\end{equation} 
where $\epsilon_{\mathbf{k}} = v_{F}^{0} |\mathbf{k}|$ is the dispersion relation corresponding to non-interacting Dirac Fermions as shown in FIG. 1 and  $|\mathbf{k}|= \sqrt{k_x^2 + k_y^2}$. 
The ground state of the non-interacting system is the filled Fermi sea and is given by,
$|F \rangle = \prod_{\mathbf{k}< k_{F}} \psi^{\dagger}_{\mathbf{k}} |0\rangle$ ,
where $|0\rangle$ is the vacuum.
\paragraph*{}
Once the interaction is switched on adiabatically the system evolves into a helical Fermi liquid (HFL) with Hamiltonian,
\begin{equation}\label{h}
H=H_0 + H_{int},
\end{equation}
where the Fermi velocity $v_{F}^{0}$ appearing in (\ref{h0}) and (\ref{h01}) is now replaced by the renormalized Fermi velocity $v_F$ of the interacting helical Fermions. The electron-electron interaction Hamiltonian $H_{int}$, written in the helical basis, is given by,
\begin{eqnarray}\label{int}
& & H_{int} = \frac{1}{2\Omega_{2D}}\int \frac{d^2 k_{1}}{(2\pi)^2}\int \frac{d^2 k_2}{(2\pi)^2}\int \frac{d^2 q}{(2\pi)^2} V(\mathbf{k_1 , k_2, q})\nonumber \\
& & \psi_{\mathbf{k_1 -q}}^{\dagger} \psi_{\mathbf{k_2 +q}}^{\dagger} \psi_{\mathbf{k_2}} \psi_{\mathbf{k_1}}, \nonumber \\
&\text{with}& \nonumber \\
& &V (\mathbf{k_1 , k_2, q}) = V(\mathbf{q}) \frac{1}{4} \Big[ 1 +  e^{i( \theta_{\mathbf{k_1}} - \theta_{\mathbf{k_{1} - q}} )} \nonumber \\ &+& e^{i(\theta_{\mathbf{k_2}} - \theta_{\mathbf{k_{2} + q}})}  +  e^{i(\theta_{\mathbf{k_1}} - \theta_{\mathbf{k_{1} - q}}+\theta_{\mathbf{k_2}} - \theta_{\mathbf{k_{2} + q}})} \Big].
\end{eqnarray}
The e-e interaction $V(\mathbf{q})$ in the above equation is such that its Fourier transform $V(|\mathbf{r-r'}|)$ is a central potential,  $\Omega_{2D}$ being the surface area of the system which I shall take as `1' for the rest of the calculations. The projected Hamiltonian corresponding to (\ref{h}) therefore, represents that of an effective spin-less Fermi liquid however, at  the cost of the electon-electron interaction becoming phase dependent \cite{LM}. In this paper, I consider a repulsive interaction, which with a sufficiently general dependence on the momentum transfer $(\mathbf{k}-\mathbf{k'})$, leads to anisotropic state with deformed Fermi surface through Pomeranchuk instability characterized by finite internal angular momentum $l$ of the corresponding order parameter. Attractive interaction usually leads to superconductivity in the BCS channel \cite{OHJ}.  The formalism considered in this paper follows closely that of the Ref. \cite{QHS}, and is sufficiently general in the sense that it depends only on some pertinent parameters derived from $V(|\mathbf{r-r'}|)$. These parameters are $V_{l}^{(n)} (k, k')$ which are determined from the following relation,
\begin{eqnarray}\label{paramet}
& &V(\mathbf{k - k'}) = \sum_{l=0}^{\infty} V_{l} (k, k') \cos l \theta_{k,k'}, \nonumber \\
& & \text{and the amplitudes $V_l$ are given by,}\nonumber \\
& & V_{l} (k, k') = \pi (1+ \delta_{l,0}) \int_{0}^{\infty} dr r V(r) J_{l} (kr) J_{l} (k' r),
\end{eqnarray}
where $k = |\mathbf{k}|$, $\theta_{k,k'} = (\theta_{k} -   \theta_{k'})$, and $J_{l} (kr)$ are the ordinary Bessel functions of first kind. The quantity $V_{l}(k, k')$ represent the amplitude of the interaction in $l$'th angular momentum channel. The symbol `$(n)$' mentioned above represents the order of the partial derivatives of $V_{l}(k, k')$ with respect to $k$ and $k'$. Such a parametrization is quite common in the literature. $V_{l}(k, k')$ and their partial derivatives of all orders constitute the microscopic parameters of the theory. 
\section{mean-field theory and microscopic expression for projected Landau parameters}\label{mft}
To describe a microscopic theory of Pomeranchuk instability on the surface of a 3D TI, I first recall the Landau functional corresponding to the effective spin-less HFL \cite{LM},
\begin{eqnarray}\label{pheno}
& & \delta \bar{E}[\delta{\bar{n}_{\mathbf{k}}}] = \int \frac{d^2 k}{(2\pi)^2} (\epsilon_{\mathbf{k}} - \mu) \delta\bar{n}_{\mathbf{k}}  \nonumber \\ 
 & + & \frac{1}{2} \int \frac{d^2 k}{(2\pi)^2} \int \frac{d^2 k'}{(2\pi)^2} \underbrace{ \sum_{l=0}^{\infty} \frac{\bar{F}_{l}}{\rho(\epsilon_F)} \cos (l\theta_{\mathbf{k,k'}})}_{\bar{f}(\mathbf{k, k'})} \delta\bar{n}_{\mathbf{k}} \delta\bar{n}_{\mathbf{k'}}, \nonumber \\
\end{eqnarray}
where $\bar{F}_{l}$ are the projected Landau parameters whose microscopic expression in terms of $V_{l} (k, k')$ shall  be derived on the Fermi surface, $\bar{f}(\mathbf{k, k'})$ is the Landau interaction function, and $\rho(\epsilon_{F}) = \frac{k_{F}}{2\pi v_{F}}$ is the density of states (DOS) at the Fermi surface. In the above equation $\delta{\bar{n}_{\mathbf{k}}} = \bar{n}_{\mathbf{k}} - \bar{n}^{0}_{\mathbf{k}} $ where $\bar{n}_{\mathbf{k}} = \langle \psi^{\dagger}_{\mathbf{k}} \psi_{\mathbf{k}}\rangle$ is the quasi-particle distribution function and the superscript `0' denotes the corresponding ground state distribution of the bare particles. The requirement of $\delta \bar{E} <0$ leads to the PI condition,
\begin{equation}\label{pi}
\bar{F}_{l} > -1
\end{equation}
applied only to the projected Landau parameters. Since a mean-field theory is equivalent to doing a one-loop static Hartree-Fock approximation (because of static interaction considered in this case) to the Fermionic self energy, one can readily write down the ground state energy,
\begin{eqnarray}\label{gdst}
\bar{E} &=& \int \frac{d^2 k}{(2\pi)^2}  \bar{n}_{\mathbf{k}} \Bigg[(\epsilon_{\mathbf{k}} - \mu)+ \frac{1}{2} \int \frac{d^2 k'}{(2\pi)^2} \nonumber \\ & & \left(V(0) - \frac{1}{2} (1+ \cos \theta_{k,k'}) V(\mathbf{k - k'}) \right) \bar{n}_{\mathbf{k'}} \Bigg],
\end{eqnarray}
where the static self energy is given by,
\begin{equation}\label{self}
\Sigma(\mathbf{k}) = - \int \frac{d^2 k'}{(2\pi)^2} \left[\frac{1}{2} (1+ \cos \theta_{k,k'}) V(\mathbf{k - k'}) -V(0)\right] \bar{n}_{\mathbf{k'}} .
\end{equation} 
The quasi-particle distribution function is given by, $\bar{n}_{\mathbf{k}} = \Theta(- \mathcal{E}_{\mathbf{k}})$ with a renormalized dispersion relation, $\mathcal{E}_{\mathbf{k}} = \epsilon_{\mathbf{k}} - \mu - \Sigma(\mathbf{k})$. With this definition of the renormalized dispersion relation the ground state energy takes the form,
\begin{equation}\label{gdst1}
\bar{E} = \int \frac{d^2 k}{(2\pi)^2} \bar{n}_{\mathbf{k}} [\mathcal{E}_{\mathbf{k}} + \frac{1}{2} \Sigma(\mathbf{k})]. \nonumber
\end{equation}
The variation of the above equation with respect to quasi-particle distribution function leads to  (\ref{pheno}) but with a renormalized dispersion relation, $\mathcal{E}_{\mathbf{k}}$ and a microscopic expression for the Landau interaction function which is given by,
\begin{eqnarray}\label{fl0}
\bar{f}(\mathbf{k, k'}) = \sum_{l=0}^{\infty} & & \frac{(2-\delta_{l,0})\bar{F}_{l}}{\rho(\epsilon_F)} \cos (l\theta_{k,k'}) = \Big[V(0) \nonumber \\ &-& \frac{1}{2} (1+ \cos \theta_{k,k'}) V(\mathbf{k - k'})\Big]
\end{eqnarray}
on the Fermi surface. In the above equation, I have used the conventions and notations used in Ref. \cite{Vignale}. Further, using (\ref{paramet}) it is easy to find the microscopic expression for the projected Landau parameters,
\begin{eqnarray}\label{fl}
\bar{F}_{l}&=& \rho(\epsilon_F) \Bigg[ \delta_{l,0} \bar{V} - \frac{V_l}{4} (1+ \delta_{l,0}) \nonumber \\ &-& \frac{1}{8} (V_{l+1} + (1+ \delta_{l,1}) 
V_{|l-1|}) \Bigg],
\end{eqnarray}
within the mean-field description where the quantity $\bar{V} = V(0) = \int d^2 r V(r)$ is the volume (2D) average of the interaction; see Appendix \ref{app00} for detailed derivation. It is due to the presence of strong SOC, the microscopic expression of the projected Landau parameter corresponding to $l$'th angular momentum channel involves interaction strength corresponding to $l, (l-1 ), \; \text{and} \; (l+1)$'th angular momentum channels. However, if a correction to the mean-field theory is considered then more higher and lower angular momentum channels are expected to appear in the Landau parameter corresponding to the $l$'th angular momentum channel. The renormalization of Fermi velocity, $v_F$ in the HFL is the analogue of the quasi-particle mass renormalization of the SU(2) invariant Fermi liquid. In absence of Galilean invariance the standard way to derive $v_F$ is to use the fact that due to adiabatic continuity the total flux of quasi-particles is equal to that of the free electrons. A straight forward calculation  (see Appendix \ref{app0}) leads to renormalized Fermi velocity expressed in terms of the microscopic parameters of the theory, and is given by,
\begin{equation}\label{vf}
v_{F}^{0} = v_{F} (1 + \bar{F}_1), 
\end{equation}
where $v_{F}^{0}$ is the Fermi velocity of the bare particles and $\bar{F}_{1} = -\frac{\rho(\epsilon_F)}{8} (2V_0 + 2V_1 +V_2)$ which is (\ref{fl}) with $l=1$. Here $V_l = V_{l} (k_F , k_F)$ which corresponds to (\ref{paramet}), evaluated at $k=k_F$ and $k' =k_F$.
\section{Small Fermi surface deformation and instability condition}\label{defo}
In order to quantify the small deformation of the Fermi surface I first expand the Fermionic self energy in circular harmonics,
\begin{equation}\label{sigm}
\Sigma(\mathbf{k}) = \sum_{m=0}^{\infty} \Sigma_{m}(k) \cos(l \theta_{\mathbf{k}}),
\end{equation}
where $m \in \mathbb{Z}_{\geq 0}$ are the angular momentum quantum numbers of the helical electron-hole pairs. The renormalized dispersion then takes the following form,
\begin{equation}\label{rmd}
\mathcal{E}_{\mathbf{k}} = \mathcal{E}_{0} (k) -  \sum_{m=1}^{\infty} \Sigma_{m}(k) \cos(m \theta_{\mathbf{k}})
\end{equation}
where $\mathcal{E}_{0}(k) = \epsilon_{\mathbf{k}} - \mu - \Sigma_{0}(k) $ is the symmetric part of the dispersion relation before the instability sets in. The symmetric part of the renormalized dispersion relation further determines the Fermi surface (FS) via the relation $\mathcal{E}_{0}(k_F)=0$, where $k_F$ remains unchanged due to Luttinger's theorem, before the instability sets in \cite{Lutt1, Lutt2}. For any small deformation of the FS corresponding to $k_F \rightarrow k_F + \delta k_F (\theta)$, with $|\delta k_{F} (\theta)| << k_{F}$ the symmetric part of the dispersion relation can be linearised, by Taylor expanding it, to have,
\begin{equation}
\mathcal{E}_{0}(k_F + \delta k_F(\theta)) = \underbrace{\mathcal{E}_{0}(k_F)}_{=0}+ v_F \delta k_F (\theta),
\end{equation}
when in the Taylor expansion in the above equation, for all $n\geq 1$, the following conditions are satisfied:
\begin{eqnarray}\label{cond}
\frac{(\delta k_F)^{n}}{(n+1)!} \left( \frac{\partial ^{n+1} \mathcal{E}_{0}(k)}{\partial k^{n+1}} \right)_{k_F}  &\ll & v_F, \;  \text{and}\nonumber \\
 \frac{(\delta k_F)^{n-1}}{n!}  \left( \frac{\partial ^{n} \Sigma_{l}(k)}{\partial k^{n}} \right)_{k_F}  & \ll & v_F,
\end{eqnarray}
not only for $l=0$ but also for all $l\neq 0$. Furthermore, with these conditions being satisfied we can safely assume $\Sigma_{l}(k_F + \delta k_F (\theta)) \approx \Sigma_{l}(k_F) = \Sigma_{l}$, near the FS for small deformation of the same. This signifies that the $|\mathbf{k}|$ dependent amplitude of the deformation potential can be approximated to be a constant near the FS. The above conditions further restricts us to consider only small FS deformations. In the symmetry-broken phase with $\tilde{k}_{F} (\theta)= k_F + \delta k_F (\theta)$, the FS is no longer determined by the symmetric part of the dispersion relation, instead it is determined by the full renormalized dispersion relation, $\mathcal{E}({k_F + \delta k_F} )= v_F \delta k_{F} (\theta_{\mathbf{k}}) -  \sum_{l=1}^{\infty} \Sigma_{l} \cos(l \theta_{\mathbf{k}}) = 0$ which implies,
\begin{equation}\label{delkf}
\delta k_F (\theta_{\mathbf{k}}) = \sum_{l=0}^{\infty} \frac{\Sigma_l}{v_F} \cos (l\theta_{\mathbf{k}}).
\end{equation}
The summation in the above equation extends from 0 to $\infty$, because in the isotropic phase corresponding to $l=0$ the deformation  $ \delta k_{F} (\theta)$ vanishes. This further implies $\Sigma_0 = 0$ from the above equation. Using (\ref{paramet}), (\ref{self}) and (\ref{sigm}) it is straight forward to construct a self consistency equation for $\Sigma_l$ for all $l\neq 0$ which is given by,
\begin{eqnarray}\label{selfcons}
& &\Sigma_l = \frac{1}{\pi} \int_{0}^{2\pi} \frac{d \theta_{k'}}{4\pi^2} \int_{k_F}^{k_F + \delta k_{F}(\theta_{k'})} k' dk' \Bigg[ \frac{\pi}{2} V_l (k_F , k') \nonumber \\ & + & \frac{\pi}{4} \lbrace V_{l+1} (k_F , k') + (1+ \delta_{l,1}) V_{|l-1|} (k_F , k') \rbrace \Bigg] \cos(l\theta_{k'}). \nonumber \\
\end{eqnarray}
It is worthwhile to note that the above equation is self-consistent via the dependence of $\delta k_{F}(\theta_k)$ on $\Sigma_l$ through (\ref{delkf}). The self-consistency equation presented above can further be reduced to a polynomial equation in $\Sigma_{l}$ to a given order in $\delta k_F(\theta_k)$ if one expands $V_l (k_F , k')$ corresponding to the above equation in a Taylor expansion,
\begin{equation}
 V_l (k_F , k') = V_l (k_F , k_F) + \sum_{j=1}^{\infty} \frac{(k'-k_F)^j}{j!} V_l^{(j)} (k_F , k_F). \nonumber
\end{equation}
 Here $V_l^{(j)} (k_F , k_F)$, being the $j$'th partial derivative of $V_l (k_F , k')$ with respect to $k'$ at $k_F$. In the mean-field theory described here $V_l (k_F , k_F) = V_l$ and $V_l^{(j)} (k_F , k_F)$ are the microscopic parameters of the system characterizing the nature of the interaction potential including its magnitude and curvature on the Fermi surface.
\paragraph*{}
Solving the self-consistency equation (\ref{selfcons}) to the lowest order in $\delta k_F (\theta)$, i.e. by taking $V_l (k_F , k') \approx V_l (k_F, k_F)$, the microscopic form of the instability equation for $l\neq 0$ can be derived to be,
\begin{equation}\label{insta}
\frac{\mathcal{V}_{l}}{V_c}=\left[ \frac{V_l}{V_c /2} + \frac{V_{l+1} + (1+\delta_{l,1} ) V_{|l-1|}}{V_c} \right] = 1,
\end{equation} 
(see Appendix \ref{app1} for detailed calculations) where the critical interaction strength required for the instability to set-in is $V_c = \frac{8}{\rho(\epsilon_{F})}$, which is determined by the DOS $\rho(\epsilon_{F})$ at the Fermi surface, and $\mathcal{V}_{l} = [2V_l + (1+\delta_{l,1} ) V_{|l-1|} + V_{l+1}]$. At this point (\ref{insta}) demands a considerably detailed analysis. Instability can occur at a particular angular momentum channel, $l$ not only if the interaction strength in that channel exceeds a value $V_c /2$ but also if the interaction strength in $l\pm 1$ channel exceed $V_c$. Therefore, one may in principle get a Fermi surface instability in $l$'th angular momentum channel by a sufficiently strong interaction in the angular momentum channel one step lower to $l$, i.e. $(l-1)$. However, the interaction in $(l-1)$'th channel must be more than twice of the same in the $l$'th channel. This finding has more interesting consequences in the case of nematic instability corresponding to $l=2$, to be discussed in Section \ref{okh}. As a whole, for $\frac{\mathcal{V}_{l}}{V_c} <1$ the un-deformed FS is a minimum energy configuration. However, for the condition, $\frac{\mathcal{V}_{l}}{V_c} >1$ the corresponding deformed FS is the minimum energy configuration, and this condition is the microscopic version of the PI condition (\ref{pi}). This can be easily seen by rewriting the above equation (\ref{insta}) using (\ref{fl}). Using the expression of the renormalized Fermi velocity it is easy to appear at following microscopic form of the PI condition (\ref{pi}),
\begin{equation}\label{l=1}
\left[ 2 V_l + V_{l+1} + (1+\delta_{l,1} ) V_{|l-1|} \right] -(2V_1 + 2V_0 + V_2)\geq \frac{16 \pi v_{F}^{0}}{k_F},
\end{equation}
for all $l= \, 1,\, 2, \, 3, ...$, where $v_{F}^{0}$ is the Fermi velocity of the bare particles and the equality sign represents the critical point where the PI sets in. The above equation represents one of the main results of the paper, viz., the microscopic expression of the PI condition.  This equation further lacks any solution for $l=1$ establishing the absence of $l=1$ PI, i.e., absence of any spontaneous current generation. It is worthwhile to mention that the critical value of the interaction strength $V_c = \frac{16 \pi v_{F}^{0}}{k_F} + (2V_0 + 2V_1 + V_2)$ depends not only on the interaction strength in all of the $l=0,\, 1,\, 2$ angular momentum channels but also on the properties of the system in absence of interaction, viz., the Fermi momentum $k_F$ and $v_{F}^{0}$. The absence of $l = 1$ PI thereby becomes intimately related to the above mentioned properties of the non-interacting system through $V_c$. Within the mean-field theory, the equation (\ref{l=1}) in turn gives a manifestation of the absence of  $l = 1$ PI only in terms of properties of the non-interacting system.
\paragraph*{}
Furthermore, from (\ref{pi}) and (\ref{fl}) it is easy to recognize that for $l=0$ the instability equation takes the form, $2V_0 - V_2 - 8\bar{V} > \frac{16 \pi v_{F}^{0}}{k_F}$. In this relation, $\bar{V}$ can be evaluated by taking the limiting value, $\lim_{\mathbf{k} \rightarrow \mathbf{k}'} V(\mathbf{k}-\mathbf{k}')$. Using (\ref{paramet}) one finds $\bar{V} = \sum_{l=0}^{\infty} V_{l}(k_F , k_F)$ on the Fermi surface. Therefore the $l=0$ instability equation takes the form,
\begin{equation}
\left[ -2V_0 - V_2 - 8\sum_{l=1}^{\infty} V_{l}(k_F , k_F) \right] > \frac{16 \pi v_{F}^{0}}{k_F},
\end{equation}
which signifies that the instability at $l=0$ channel (corresponding to a phase separation instability) can be achieved only via attractive interaction in not only the $l=0$ channel but also in any other angular momentum channels. This is the microscopic manifestation of the fact that unlike the SU(2)-FL, in HFL an instability towards the phase separation can be achieved not only through attractive density-density interaction but also as a result of spin-charge or even pure spin-spin interaction \cite{LM}. However, it is important to note that the interaction must be attractive in the angular momentum channels which are participating in the instability. This will be further concretized in section \ref{tpt} with the example of a $\delta$-shell potential.
\section{Nature of the phase transition}\label{pt}
In order to investigate the nature of the phase transition emerging through PIs, I now solve the self consistency equation in next higher order in $\delta k_{F}(\theta)$, i.e., use $V_{l}(k_F , k') = V_{l}(k_F , k_F) + V_{l}^{(1)} (k_F , k_F) \delta k_{F} (\theta_k)$. Using (\ref{delkf}) the following self-consistency equation can be found for all $l \geq 2$ (since $l=1$ instability is excluded),
\begin{equation}\label{nh1}
\Sigma_{l} = \frac{1}{8 \pi} \left[ \frac{\mathcal{V}_l k_F}{v_F} \Sigma_l + \frac{\mathcal{V}_{l}^{(1)}}{4 v_F^{3}} \Sigma_{l}^{3} \right],
\end{equation}
where $\mathcal{V}_{l}$ has been defined earlier (see discussions near (\ref{insta}), with $l \neq 1$) and $\mathcal{V}_{l}^{(1)} = 2V_{l}^{(1)} + V_{l-1}^{(1)} + V_{l+1}^{(1)}$ (see Appendix \ref{app1} for the details of the calculations). This equation has two solutions, one corresponding to $\Sigma_{l} = 0$ which represents a phase corresponding to zero deformation of the Fermi surface which minimizes the energy when the interaction strength is less than $V_c$. The other solution is,
\begin{equation}\label{odpar}
\frac{\Sigma_l}{k_F v_F} = \pm \sqrt{-\frac{V_c}{k_F \mathcal{V}_{l}^{(1)}}} \left( \frac{2V_{l} + V_{l-1} + V_{l+1}}{V_c} -1\right)^{1/2},
\end{equation}
which describes the physical Fermi surface deformation amplitude and its variation with respect to interaction strength if $\mathcal{V}_{l}^{(1)}<0$ (otherwise the term within the square root would have been imaginary!), and the corresponding configuration minimizes the energy. This situation is depicted in FIG 2(a) and is structurally quite similar to that of Ref. \cite{QHS}, however, bears remarkable qualitative differences only because of the presence of strong SOC on the surface of a 3D TI. 
\begin{figure}[h!]\label{defor}
\includegraphics[scale=0.49]{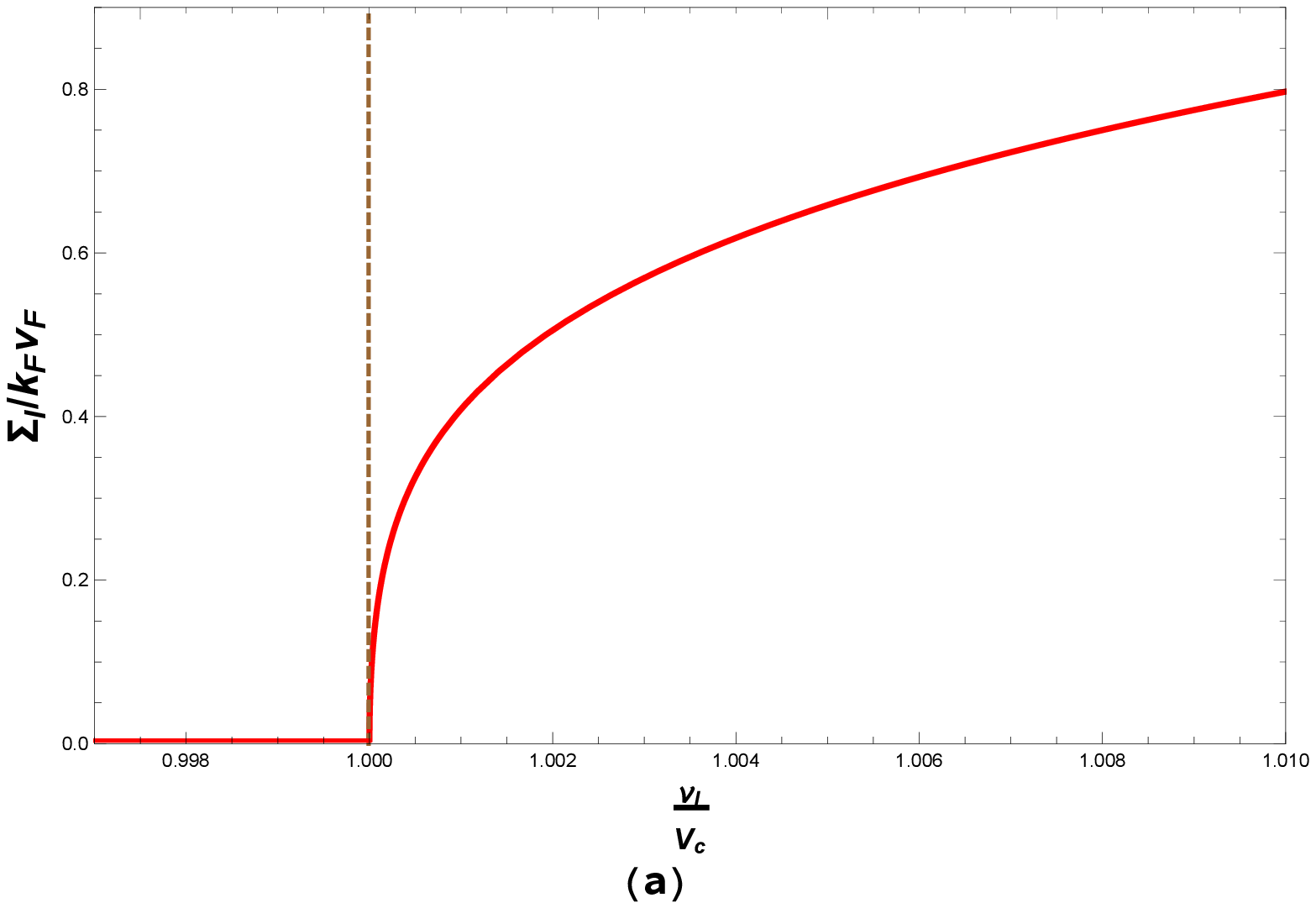} 
\includegraphics[scale=0.49]{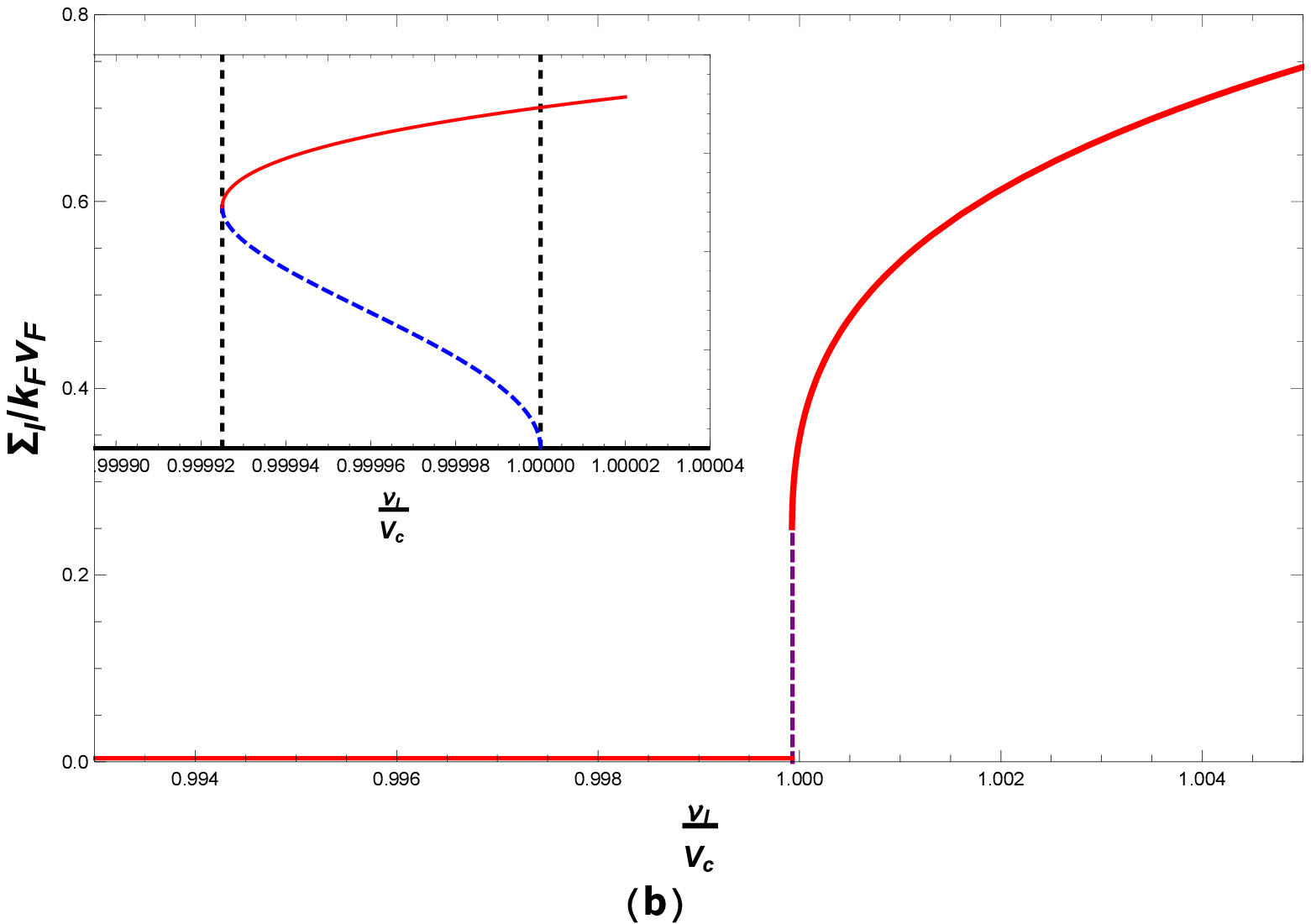} 
\caption{(a) Plot of the amplitude of the Fermi surface deformation $\frac{\Sigma_l}{k_F v_F}$ as a function of interaction strength. Here $\mathcal{V}_{l} = 2V_{l} + V_{l-1} + V_{l+1}$, and $-\frac{V_c}{k_F \mathcal{V}_{l}^{(1)}} =1$ on the Fermi surface. The black vertical line indicates the critical interaction strength $V_c$. (b) Plot of the solution of (\ref{insta_1}) with $v_{l}^{(1)} + \frac{v_{l}^{(2)}}{2} = -|\eta|= -1/100$ and $v_{l}^{(3)} =-1$. In this plot, the vertical dashed line at the critical point represents a first order jump in the order parameter. In the inset the bound for the interaction strength represented by thick vertical dashed lines is plotted (see text for explanation).}
\end{figure}
For $\mathcal{V}_{l}^{(1)}<0$, the above equation further describes the appearance of a broken-symmetry phase through PI with a mean-field order parameter $\langle \Psi_{l} \rangle = \sum_{\mathbf{k}} \cos(l \theta_{k}) \Theta (k_F + \delta k_{F}(\theta_{k}) - \mathbf{k})$, corresponding to $l$'th angular momentum channel, and a second order quantum phase transition with a mean-field critical exponent $1/2$, as expected. This mean-field order parameter is indeed proportional to the amplitude of the Fermi surface deformation $\frac{\Sigma_l}{k_F v_F}$ thorough $\delta k_F(\theta_{k})$, and is given by
\begin{equation}\label{piod}
\langle \Psi_{l} \rangle = \frac{k_F}{2\pi} \frac{\Sigma_l}{k_F v_F}.
\end{equation}
This will further be discussed in detail in the case of the nematic instability in the next section. In situations when the interactions in the angular momentum channels exceed the critical interaction strength in a hierarchical manner from smaller values of $l$ to larger ones (say a ``\textit{hierarchy assumption}"), even a sufficiently  strong interaction in $(l-1)$ channel in principle can drive the system towards a PI in the $l$'th channel. This situation is described as phase diagrams in FIG 3.
\begin{figure}[h!]\label{phased}
\includegraphics[scale=0.45]{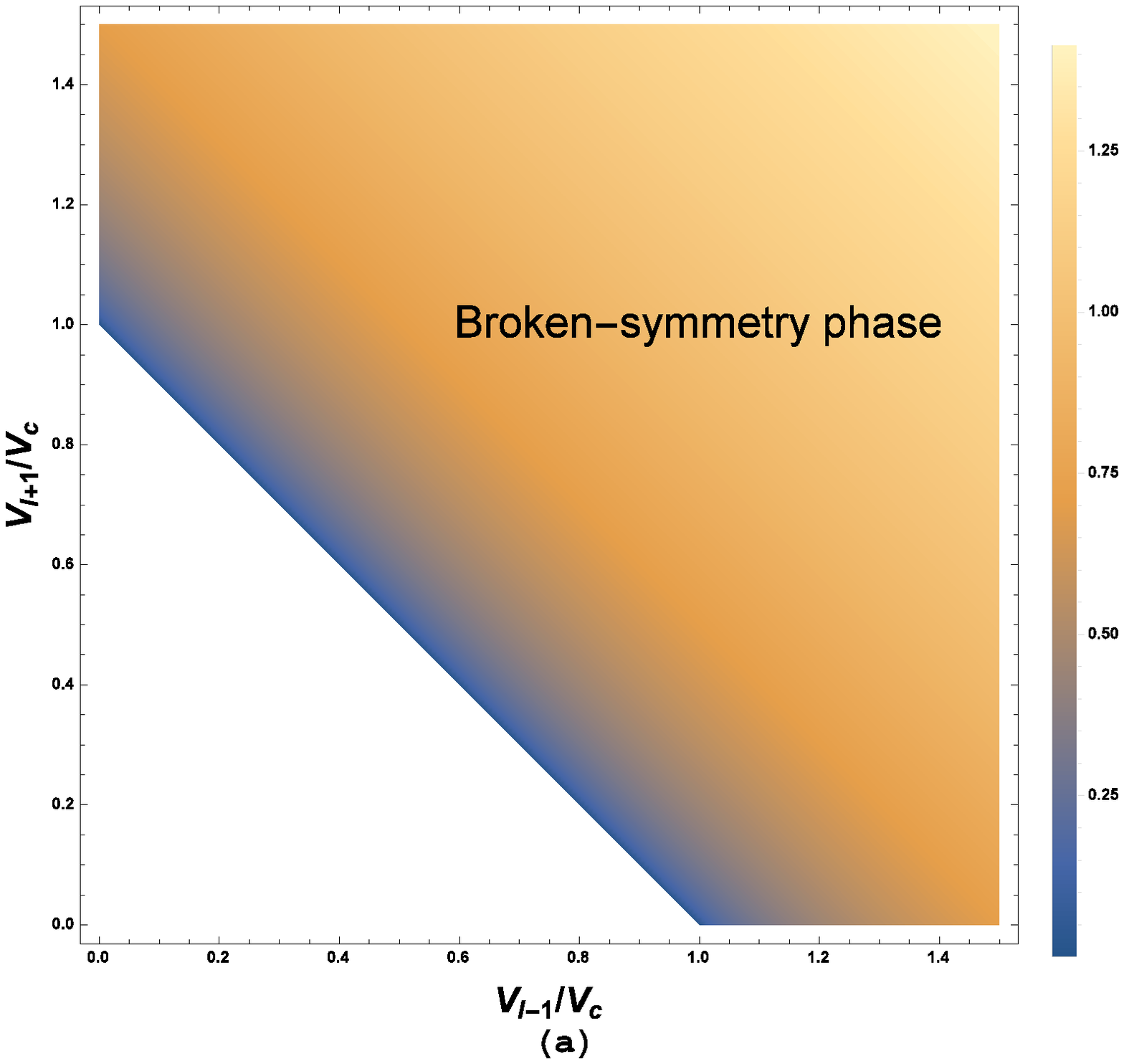}
\includegraphics[scale=0.45]{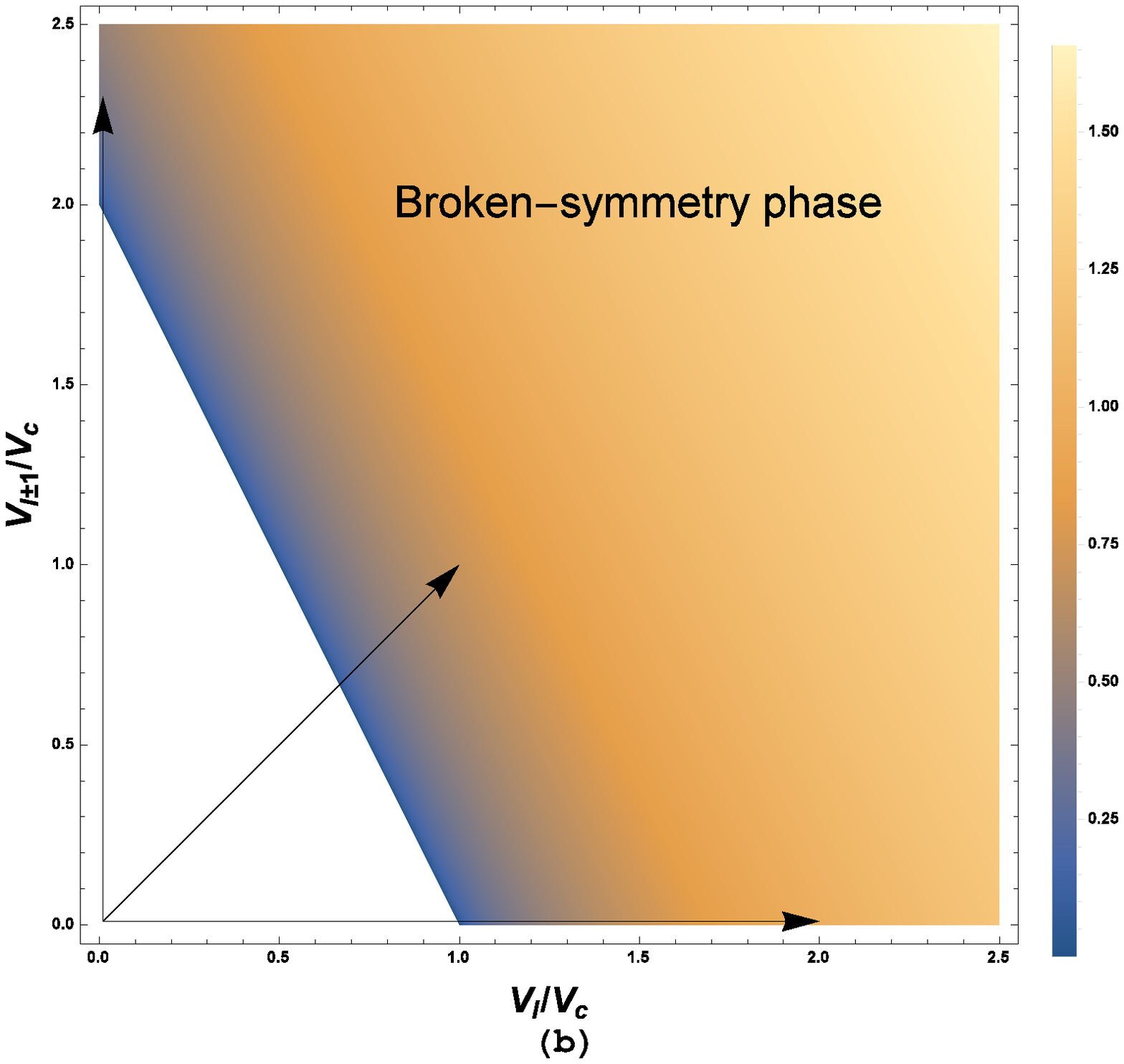}
\caption{The white portion of the diagram represents the isotropic phase; symmetry broken phase is indicated in the diagram. Density represents the amplitude of the deformation corresponding to (\ref{odpar}). (a) This phase diagram corresponds to the situation when the interaction in $l$'th angular momentum channel is zero. In this case, the critical strength required for instability to occur is the same for both $(l-1)$'th and $(l+1)$'th channels. This further represents the situation when PI corresponding to $l$'th channel can be achieved through interaction in the $l$'th channel exceeds the critical strength. (b) This phase diagram corresponds to the situation when the interaction in $l$'th angular momentum channel and $(l\pm 1)$'th channel compete with each other. The horizontal black arrow along the $V_l /V_c$-axis represents the situation when interaction in $(l\pm1)$'th channel is zero. In this case the critical strength required is $V_c / 2$.  The vertical black arrow represents the situation when $V_l = 0$ and the instability can be achieved in either $(l-1)$'th or $(l+1)$'th channel. In the case when all the interactions are present, the arrow would be in the ($V_l/V_c - V_{l\pm1}/V_c $)- plane (corresponding arrow is shown in the diagram).}
\end{figure}
 In the case when $V_l = \eta \frac{V_c}{2}$, with $0<\eta<1$, the PI condition corresponding to (\ref{insta}) takes the form $(V_{l+1} + V_{l-1})>(1-\eta)V_c$, a situation depicted in FIG 3(a) which corresponds to $\eta = 0$. In the opposite case when $\eta \geq 1$ the PI in the $l$'th channel appears irrespective of the interaction strength in the other available channels (in this case $V_{l-1}$ and $V_{l+1}$, within mean-field theory).  The case corresponding to arbitrary $\eta$ is described in FIG. 3(b) where all the three angular momentum of interaction compete with each other.
\paragraph*{}
When $\mathcal{V}_{l}^{(1)}>0$, the amplitude of the Fermi surface deformation corresponding to (\ref{odpar}) turns out to be imaginary and therefore unphysical. In this case the self-consistency equation is needed to be solved in the next higher order in $\delta k_F (\theta_k)$ \cite{QHS}. For the ease of notation I now introduce few dimensionless parameters, $\lambda_{l} = \frac{\Sigma_l}{k_F v_F}$, $v_{l}^{(n)}= \frac{\mathcal{V}_{l}^{(n)}k_{F}^{n}}{V_c}$, and $v_{l} = \frac{V_l}{V_c}$, following Ref. \cite{QHS}. The self-consistency equation corresponding to (\ref{odpar}) now takes the form (see Appendix \ref{app1} for details),
\begin{equation}\label{insta_1}
\frac{v_{l}^{(3)}}{48} \lambda_{l}^{5} + \frac{1}{4}\left( v_{l}^{(1)} + \frac{v_{l}^{(2)}}{2} \right) \lambda_{l}^{3} + (v_{l} -1) \lambda_{l} = 0,
\end{equation}
which has one solution, $\lambda_{l} = 0$ corresponding to the isotropic phase with no deformation. As a characteristic of the mean-field theory, this equation is structurally the same as that obtained in the case of SU(2) invariant Fermi liquid and therefore all the conclusions corresponding to the latter hold true \cite{QHS}. However, the qualitative differences originate from the strong SOC present in the system. For completeness the situations are described here too. When $v_{l}^{(3)} < 0$, one gets two types of solutions, one corresponding to $v_{l}^{(1)} + \frac{v_{l}^{(2)}}{2} < 0$, and the other corresponding to $v_{l}^{(1)} + \frac{v_{l}^{(2)}}{2}>0$.
\paragraph*{}
 For $v_{l}^{(1)} + \frac{v_{l}^{(2)}}{2} = -|\xi |$, the solution of the above equation grows in a second-order fashion as shown in the FIG 2(a) however, with a shifted quantum critical point given by, $\mathcal{V}_l / V_c = (1-3\xi ^2/4)$. Therefore, the higher order correction in $\delta k_F$ has its effect in shifting the quantum critical point with the shift being controlled by the nature of the central interaction and its curvature on the Fermi surface. In the other case, for $v_{l}^{(1)} + \frac{v_{l}^{(2)}}{2} = |\xi|$ the behaviour of the solution is plotted in FIG 2(b), which shows a jump in the order parameter corresponding to a first order phase transition, before the PI sets in. Analysing the solution it is easy to find a bound in the interaction strength which is given by,
 \begin{equation}
 \left( 1 - \frac{3 \xi ^2}{4} \right) \leq \frac{\mathcal{V}_{jump}}{V_c} < 1.
 \end{equation}
The qualitative difference from the SU(2)-FL once again appears in the above equation in the form of the fact that the interaction corresponding to the jump depends on the interactions in $l$, $(l-1)$, and $(l+1)$'th angular momentum channels. Therefore the conclusions corresponding to the ``\textit{hierarchy assumption}" introduced earlier remain valid. From the inset of FIG 2(b), it is easy to see that initially the order parameter develops even if the interaction strength decreases however, such a decrease is bounded from below as indicated in the equation above. On reaching the lower bound the order parameter starts growing in a second order fashion and PI sets in when the interaction strength exceeds $V_c$. Therefore, in this case the system goes through a weak first order phase transition before it ultimately enters in to the broken-symmetry phase. First order transition here is argued to be weak because it appears only in a very narrow parameter regime.
\paragraph*{}
For $v_{l}^{(3)} < 0$, one need to consider solving the self-consistency equation to even higher order in $\delta k_{F} (\theta)$ to resolve the issue of unphysical solutions. However, since only the small Fermi surface deformations are considered here, appearance of such parameter ranges corresponding to $v_{l}^{(3)} < 0$ are restricted. 
\paragraph*{}
It is worthwhile to point out that, in the expression of the Pomeranchuk order parameter corresponding to (\ref{piod}) the renormalized Fermi velocity $v_F$ appears in the denominator. The renormalized Fermi velocity depends on the interaction strengths in $l=0,\;1,\;2$ angular momentum channels. However, since $l=1$ PI is ruled out $v_F$ can not diverge even if the interaction strengths in these angular momentum channels exceeds the corresponding critical value, but interestingly enough, in this situation the $l=2$ PI sets in, as explained in the next section.
\section{Quadrupolar instability}\label{okh}
With sufficiently strong e-e interaction in the $l=2$ angular momentum channel the the ground state energy of an HFL is further lowered by spontaneous quadrupolar deformation of the corresponding Fermi surface, leading to a nematic phase \cite{LYM}. Such spontaneous rotational symmetry breaking in the surface states of 3D-TI has been studied considering the following effective Hamiltonian,
\begin{equation}\label{nem_r}
H= H_0 + \frac{1}{4} \int d \mathbf{r} d \mathbf{r'} \mathcal{F}(\mathbf{r-r'}) \text{Tr}[ {Q}^{\dagger} (\mathbf{r}) {Q}(\mathbf{r'})]
\end{equation}
where ${Q}(\mathbf{r})$ is a matrix of rank two which is real, symmetric, and traceless \cite{LYM}. In the projected space corresponding to $\mu > 0$, the elements of the matrix ${Q}(\mathbf{r})$ are given by, $Q_{ab}(\mathbf{r}) = \psi^{\dagger}(\mathbf{r}) [\hat{\partial}_{a} \hat{\partial}_{b} - \frac{1}{2}\delta_{ab} \hat{\partial}^{2}] \psi(\mathbf{r})$. These matrix elements determine the order parameters in the nematic phase of the HFL on the surface of a 3D-TI, and written in the original un-projected basis such order parameters contain both spin and charge degrees of freedom and linear in electrons' momentum, unlike the case corresponding to SU(2) FL \cite{OFK, LYM}. This is because, in the presence of a strong SOC the rotational symmetry involves simultaneous rotations in momentum and spin spaces generated by the total angular momentum \cite{LYM}.
\paragraph*{}
In order to understand the connection between theories based on the above Hamiltonian and the Hamiltonian (\ref{h}) with central interaction, following Ref. \cite{QHS} I need to construct a Hamiltonian whose interaction term is constrained to contain only terms of the form of (\ref{nem_r}). To demonstrate explicitly I first make a Fourier transform of the above Hamiltonian (\ref{nem_r}) so that,
\begin{widetext}
\begin{eqnarray}\label{nem_q}
H = \int \frac{d^2 k}{(2\pi)^2} (v_F k - \mu) \psi^{\dagger}_{\mathbf{k}} \psi_{\mathbf{k}} + \frac{(2\pi)^4}{8 k_{F}^{4}}  \int \left[ \frac{d^2 k}{(2\pi)^2}  \frac{d^2 k'}{(2\pi)^2}  \frac{d^2 q}{(2\pi)^2}\right] \mathcal{F}(\mathbf{q}) \bigg|\mathbf{k'} - \frac{\mathbf{q}}{2} \bigg|^2 & &  \bigg|\mathbf{k} + \frac{\mathbf{q}}{2} \bigg|^2\cos (2\theta_{\mathbf{k'} - \mathbf{q}/2} - 2\theta_{\mathbf{k} + \mathbf{q}/2}) \nonumber \\ 
& & \psi^{\dagger}_{\mathbf{k'} + \frac{\mathbf{q}}{2}} \psi_{\mathbf{k'} - \frac{\mathbf{q}}{2}} \psi^{\dagger}_{\mathbf{k} - \frac{\mathbf{q}}{2}} \psi^{}_{\mathbf{k} + \frac{\mathbf{q}}{2}},
\end{eqnarray}
\end{widetext}
where the quantity $\mathcal{F}(\mathbf{q}) = (2 \pi)^2 \mathcal{F}_{2} \delta(\mathbf{q}-0)$ and for nematic instability the value of $\mathcal{F}_2$ is usually taken to be negative \cite{LYM}. With this choice I now compare the self energy of the above Hamiltonian with the coefficient of the $\cos (2 \theta_{\mathbf{k k'}})$ of the Hartree-Fock self energy corresponding to (\ref{self}) by using the parametrization (\ref{paramet}). Simple calculation the leads to,
\begin{eqnarray}\label{nem_v}
\mathcal{V}_2 (k, k') &=& \left[ 2 V_{2}(k, k') + V_{3}(k, k') + V_{1}(k, k') \right] \nonumber \\ &=& -\frac{(2\pi)^4}{ k_{F}^{4}} k^2 k'^2 \mathcal{F}_2
\end{eqnarray}
for attractive interaction, i.e., $\mathcal{F}_2  \rightarrow -\mathcal{F}_2$. Therefore, although nematic instability originates from a quadrapolar interaction which is not a central interaction, it can be considered to be originated from a hypothetical interaction whose relevant term containing $\cos (2 \theta_{\mathbf{k k'}})$ has the from (\ref{nem_v}). This equation further suggests that the nematic instability originating from such hypothetical interaction can be achieved through any one or all of the angular momentum channels of interaction with $l= 1,2,\; \text{and} \; 3$. Moreover, on the Fermi surface the above equation  (\ref{nem_v}) describes the PI condition, $(2\pi)^4 \mathcal{F}_2 \geq -V_c$, i.e., with sufficiently large but negative values of $\mathcal{F}_2$ the system enters into the nematic phase where the equality represents the quantum critical point corresponding to the onset of PI. However, not all the angular momentum channels require equally large interaction strength to achieve the instability for example, strength required in $l=2$ channel is half of that required in $l=1$ channel. Now if I consider a situation, which is quite natural, when the angular momentum channels are activated in a hierarchical manner from the lower to higher values of $l$ (the ``\textit{hierarchy assumption}" as mentioned earlier), it is very much possible to achieve the instability only through $l=1$ channel of interaction although it is easier in $l=2$ channel. Furthermore, since $l=1$ PI is absent, the nematic instability is competed only by the instability in $l=3$ angular momentum channel, and the ``\textit{hierarchy assumption}" therefore, has its implication in this case in avoiding competing instability coming from $l=3$ angular momentum channel. 
\paragraph*{}
From (\ref{nem_v}) it is easy to recognize that $\mathcal{V}_2^{\prime} (k_F, k_F) = -\frac{(2\pi)^4 \mathcal{F}_2}{k_F} < 0$ which implies that the isotropic to nematic transition is a second order phase transition, in agreement with the Ref. \cite{LYM}. It is worthwhile to point out that in the case of the conventional Landau Fermi liquid, it has been essential to incorporate non-linear terms in the non-interacting dispersion relation to stabilize the nematic quantum critical point \cite{OFK}. However, in case of HFL which is strictly valid for finite doping only, $\mu > 0$, the linear dispersion relation corresponding to a single Dirac cone has been sufficient to stabilize the quantum critical point and phase transition becomes continuous in nature even at $T=0$ \cite{LYM}. 
\paragraph*{}
To find the relation between the nematic order parameters developed from the quadrupolar interaction corresponding to (\ref{nem_r}) and the Pomeranchuk order parameter developed from the central interaction corresponding to (\ref{h}), I now derive the nematic order parameters corresponding to the quadrupolar interaction which are defined by,
\begin{eqnarray}
& &\langle Q_{xx} (\mathbf{q}=0) \rangle = - \langle Q_{yy} (\mathbf{q}=0) \rangle  \nonumber \\ &=& \int \frac{d^2 k}{(2\pi)^2} \frac{k^2}{2 k_{F}^{2}} \cos (2 \theta_k) \Theta(k_F + \delta k_F (\theta_k) - \mathbf{k}) \nonumber \\ &\text{and}& \nonumber \\ & & \langle Q_{xy} (\mathbf{q}=0) \rangle =  \langle Q_{yx} (\mathbf{q}=0) \rangle  \nonumber \\ &=& \int \frac{d^2 k}{(2\pi)^2} \frac{k^2}{2 k_{F}^{2}} \sin (2 \theta_k) \Theta(k_F + \delta k_F (\theta_k) - \mathbf{k}) \nonumber \\
\end{eqnarray}
at zero temperature. In the isotropic phase the above order parameters vanish. Evaluating the integrals by using the limit of $k$-integrations from $0$ to $k_F +  \delta k_F (\theta_k)$ and using (\ref{delkf}) it is easy to find,
\begin{equation}\label{nem_od}
\langle Q_{xx} (\mathbf{q}=0) \rangle = - \langle Q_{yy} (\mathbf{q}=0) \rangle = \frac{k_F}{8 \pi} \frac{\Sigma_2}{v_F k_F}= \frac{\langle \Psi_2 \rangle}{4}
\end{equation}\\
and $\langle Q_{xy} (\mathbf{q}=0) \rangle =  \langle Q_{yx} (\mathbf{q}=0) \rangle = 0$, where $\Sigma_2$ is given by,
\begin{equation}
\frac{\Sigma_2}{k_F v_F} = \pm \sqrt{\frac{(2\pi)^4 V_c}{ \mathcal{F}_{2}}} \left( \frac{2V_{2} + V_{1} + V_{3}}{V_c} -1\right)^{1/2},
\end{equation}
 and (\ref{piod}) has also been used. The above equations therefore, express the nematic order parameter of a nematic helical fermi liquid on the surface of a 3D TI in terms of the Pomeranchuk order parameter obtained in the HFL description with central interaction. Nematic order parameter also grows in a second-order fashion with universal mean-field critical exponent $1/2$.
\section{$\delta$-shell model and possibility of topological phase transitions}\label{tpt}
The phase separation instability corresponding to $l=0$ angular momentum channel can not occur when the interaction is repulsive, as explain in section \ref{defo}. In this section this fact is further explained by studying a specific form of the interaction,
\begin{equation}\label{del_pot}
V(r) = g \delta^{(1)} (|\mathbf{r} - r_0|),
\end{equation}
where $g$ is the strength of the e-e interaction. This interaction profile exhibits a very sharp peak at a particular inter-particle distance $r_0$. This further becomes a contact interaction when $r_0 = 0$, while finite values of the $r_0$ represent a finite range interaction.
\paragraph*{}
Such an interaction has been studied in the case of conventional Fermi liquid in three dimensions to establish the fact that in order to have PI in isotropic systems the range of the interaction must be finite and $r_0 \sim k_{F}^{-1}$ \cite{QS}. Furthermore, study of such an interaction also have revealed that apart from PI, the three dimensional isotropic Fermi liquid also exhibit an interaction driven Lifshitz-like transition which does not involve any symmetry breaking \cite{QS}.  
Motivated by this, I study the fate of PIs in different channels and possibility of such interaction driven topological Lifshitz transition in the HFL by considering the above mentioned form of the interaction. Both repulsive ($g>0$) and attractive ($g<0$) interactions are considered keeping in mind the fact that only repulsive interaction leads to PI. 
\paragraph*{}
The PI equation corresponding to $l=0$ angular momentum channel is given by,
\begin{equation}\label{l=0}
2V_0 -V_2 - 8\bar{V} >\frac{16 \pi v_{F}^{0}}{k_F},
\end{equation} 
where $V_0 = 2\pi g r_0 [J_{0}(k_F r)]^2$, $V_2 = \pi g r_0 [J_{2}(k_F r)]^2$, and $\bar{V} = 2\pi r_0 g$. In FIG. 4 the corresponding phase is plotted by solving the above equation and it can be easily seen that the phase separation instability corresponding to $l=0$ channel can only appear when the interaction is attractive, i.e., $\frac{g}{v_{F}^{0}} < 0$. The dimensionless coupling constant $\frac{g}{v_{F}^{0}}$ and the dimensionless range of interaction $k_F r_0$ are the only two parameters available in the theory. 
\begin{figure}[h]
\includegraphics[scale=.44]{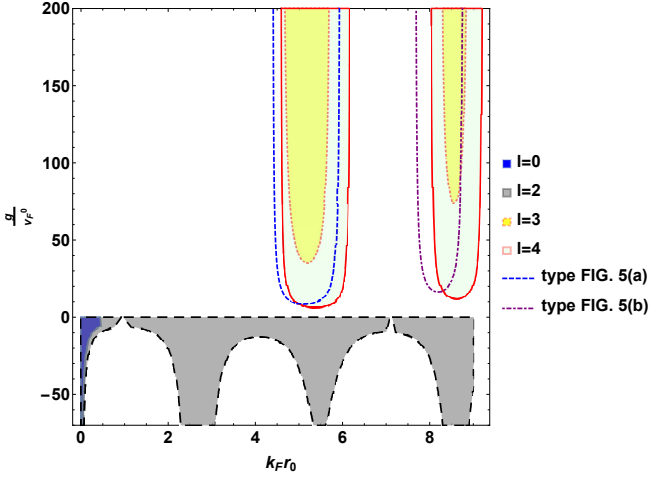}
\caption{Phase diagram showing zero-temperature instabilities for both repulsive and attractive $\delta$-shell model. $\frac{g}{v_{F}^{0}}$ is the dimensionless coupling constant and $k_F r_0$ is the dimensionless range of interaction. Attractive interaction corresponds to the negative values of $\frac{g}{v_{F}^{0}}$ and repulsive interaction corresponds to the positive values of the same. Instabilities at different angular momentum channels are indicated in the figure where the shaded regions are the symmetry broken phases. Regions bounded by long-dashed line (grey-shaded), dotted line, and solid line are symmetry-broken phase achieved in $l=2, \, 3, \, 4$ channels respectively. Regions bounded by blue short-dashed line and purple dotted-dashed line are instabilities corresponding to FIG. 5(a) and FIG. 5(b) respectively.}
\end{figure}
Similarly, the instability equation for $l \geq 2$ channels, given by (\ref{l=1}), can be solved and the corresponding phases are plotted too in FIG. 4. It turns out that there is no $l=2$ PI when the interaction is repulsive as can be seen from FIG. 4. Therefore, even within the model of the $\delta$-shell interaction the nematic instability in $l=2$ angular momentum channel can appear only for attractive interaction which is consistent with the results obtained in section \ref{okh}. Surprisingly enough, the dome corresponding to $l=3$ PI is fully contained within the dome corresponding to $l=4$ PI as can be seen from FIG. 4. Furthermore, the strength of interaction needed to achieve $l=3$ instability is much larger than that needed to achieve the $l=4$ instability which indicates that $l=4$ instability is more prone to appear. Because of the presence of strong SOC, interaction in different angular momentum channels contribute quite distinctly when instabilities in $l=3$ and $l=4$ channels are compared. This can be seen from equation (\ref{l=1}). Therefore, it turns out that the strong SOC prefers the $l=4$ instability over $l=3$ instability. However, in certain parameter regimes, mostly when the interaction is much stronger (for example when $\frac{g}{v_{F}^{0}} \sim 100$, and $k_F r_0 \sim 5$ corresponding to FIG. 4), the $l=3$ and $l=4$ PI do indeed compete, as is evident from the corresponding overlapping PI domes in FIG. 4. Moreover, PI corresponding to $l\geq 3$ for repulsive interaction can only appear if the interaction possesses a finite range, and PI domes appear only within certain values of interaction range, for example the $l=4$ PI appears for $5 \lesssim k_F r_0 \lesssim 6.25$. However, in the case of contact interaction corresponding to $r_0 = 0$, no PI can appear whatever may be the value of the strength of the repulsive interaction ($g>0$), as can be seen from FIG. 4 which doesn't exhibit any PI dome near $k_F r_0 = 0$. 
\begin{figure}[h!]
\subfloat[]{\includegraphics[width=1.0\linewidth]{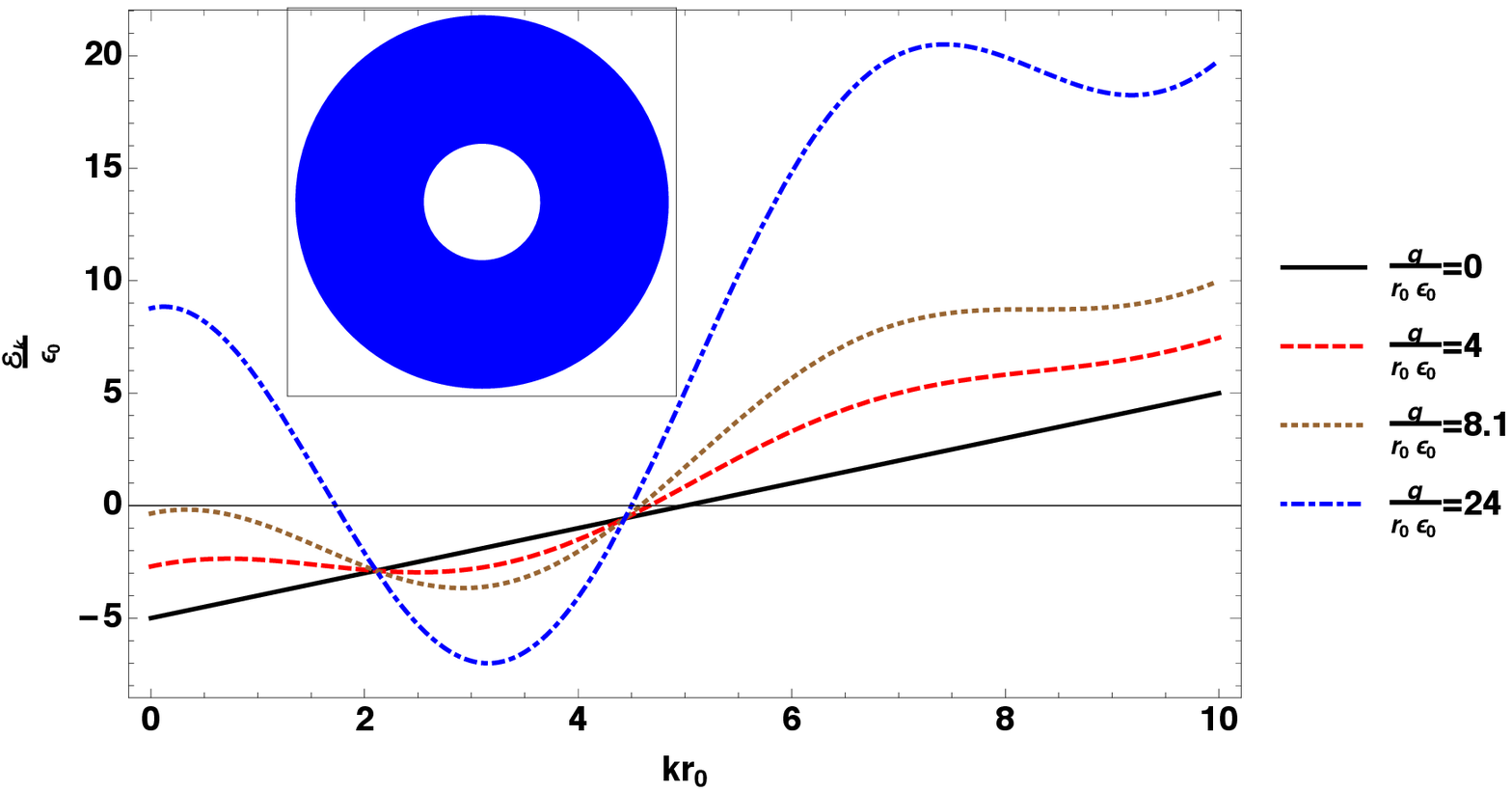}}\\
\subfloat[]{\includegraphics[width=1.0\linewidth]{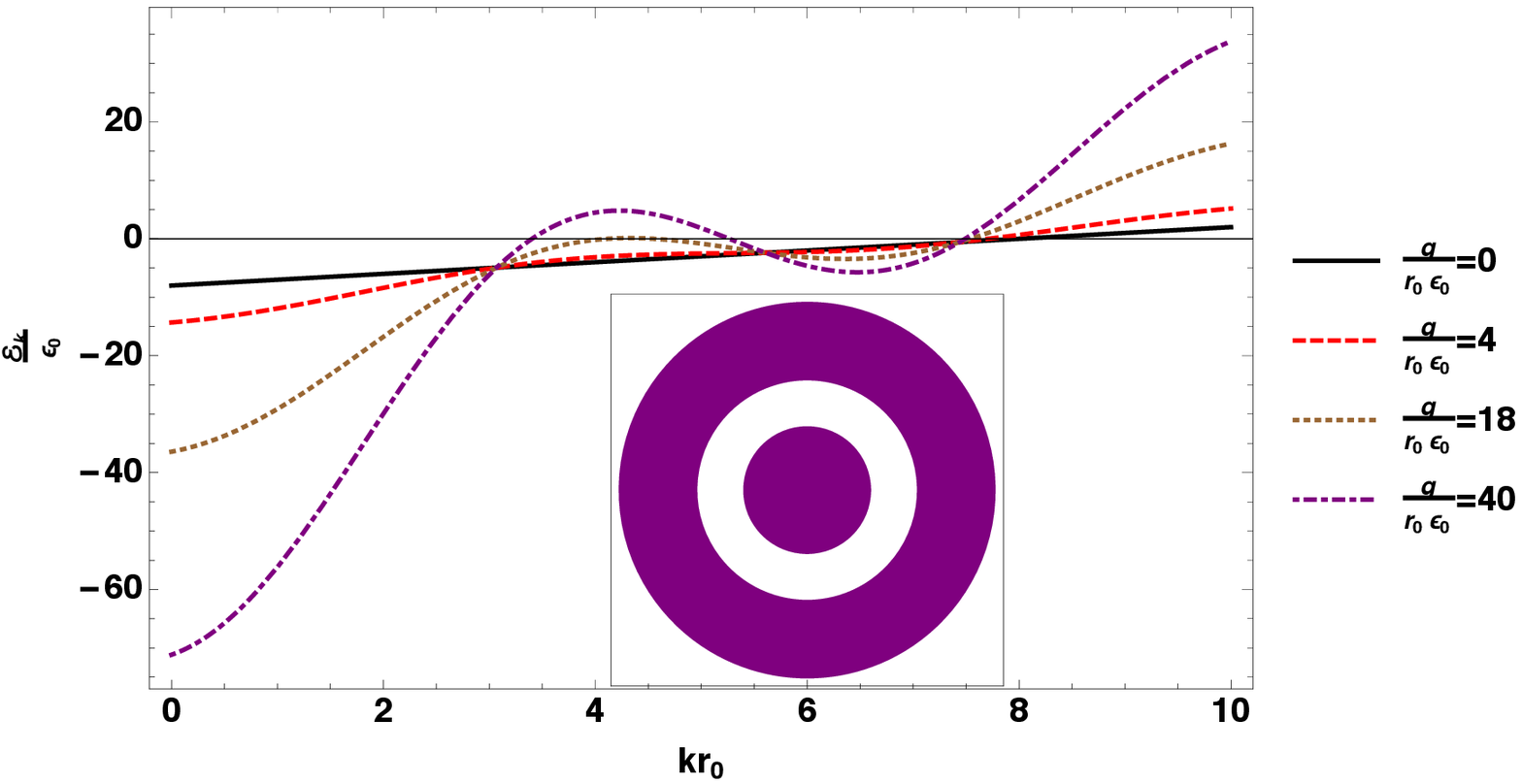}}
\caption{Renormalized dispersion relation for the $\delta$-shell model for fixed $k_F r_0$ in the Hartree-Fock ground state, (a) $k_F r_0 = 5$, (b) $k_F r_0 = 8$, where in the inset the corresponding Fermi surface configurations are drawn schematically and in the both the cases the dispersion relations are indicated in dotted-dashed curves.}
\end{figure}
\paragraph*{}
Apart from the PI, the $\delta$-shell potential also exhibits another type of Fermi surface instability which does not involve any symmetry breaking resulting from the shape deformation of the Fermi surface. This type of instability is purely an interaction driven one as explained below. To analyse such an instability, the  renormalized dispersion relation is needed to be evaluated for the interaction under consideration. The renormalized dispersion relation is given by,
\begin{equation}\label{ren_disp1}
\mathcal{E}_{\mathbf{k}} = v_{F} (k - k_F) - \Sigma (\mathbf{k}),
\end{equation}
where the mean-field self energy is given by (\ref{self}). In the state $\bar{n}_{\mathbf{k}} = \Theta (k_F - |\mathbf{k}|)$, it is easy to evaluate the renormalized dispersion for the above $\delta$-shell interaction by using (\ref{paramet}). The renormalized dispersion relation can be evaluated to be,
\begin{eqnarray}\label{ren_disp2}
 \mathcal{E}_{\mathbf{k}} &=& v_{F} (k - k_F) + k_F \int_{0}^{\infty} dr \, V(r) \Big[ J_{0} (k_F r) J_{1} (k_F r) \nonumber \\ &-& \frac{1}{2} J_{0} (k r) J_{1} (k_F r) - \frac{\pi}{16} J_{1} (k r) J_{1} (k_F r) \mathbf{H}_{0}(k_F r) \nonumber \\ &+& \frac{\pi}{16} J_{1} (k r) J_{0} (k_F r) \mathbf{H}_{1}(k_F r) \Big],
\end{eqnarray}
corresponding to an interaction $V(r)$ where $\mathbf{H}_{n} (kr)$ is the Struve function order $n$ which is the solution of the non-homogeneous Bessel differential equation \cite{AS}. Using  (\ref{vf}) and (\ref{del_pot}), the above mentioned renormalized dispersion relation is plotted as a function of dimensionless range $kr_0$ in FIG. 5, where $\epsilon_0 = \frac{v_{F}^{0}}{r_0}$, and $\frac{g}{r_0 \epsilon_{0}} \left( =\frac{g}{v_{F}^{0}} \right)$ is the dimensionless coupling as mentioned earlier. From FIG. 5 it is easy to recognize that two situations appear, (1) the dispersion relation goes above the Fermi level (corresponding to $\mathcal{E}_{\mathbf{k}}=0$) at the centre of the Fermi surface (FIG. 5(a)), and (2) the dispersion relation peaks above the Fermi level at some intermediate point $0<k<k_F$ (FIG. 5(b)). In the former, there appears a region in the $k$-space within the circular Fermi surface which is vacated. In the latter case, there exists a thin shell of vacated states for some $k$-values within the Fermi surface. The associated instabilities are completely different from the PI, as it involves a change in the occupation number away from the Fermi surface and do not involve any symmetry breaking. However, the topology of the Fermi surface changes. Therefore, the phase transition occurring in these cases can be considered to be interaction driven topological phase transition similar to the Lifshitz transtion, in analogy with the corresponding SU(2)-FL in three dimensions (3D) \cite{QS, Vol, Lifsh}. It is worthwhile to mention that the phase transition corresponding to FIG. 5(a) and (b) appear only when $k_F r_0 \sim 5$  and $k_F r_0 \sim 8$ respectively, not at any other values of the range of the interaction as can be seen from FIG. 4. Furthermore, this type of topological phase transition compete with the Pomeranchuk type phase transitions too, which is easily recognized from the overlapping of the corresponding domes (see FIG. 4). However, such a competition is exhibited only between the topological FS instability and $l=3,4$ PIs. When the range of interaction is considered to be longer (the value of $k_F r_0 > 10$ in FIG. 5), the PI domes reappear but are not competed by the topological instabilities.
\paragraph*{}
It is worthwhile to point out that a more realistic hardcore potential of the form $V(r) = g k_F \Theta(r_0 - |\mathbf{r}|)$ does not exhibit any PI, however, exhibit only one type of interaction-driven Lifshitz transition corresponding to the appearance of an annular sheet of hole Fermi sea within the $k_F$. This can be obtained by solving the PI equation and equation (\ref{ren_disp2}) numerically. The analysis performed in this section suggests that the interaction-driven topological Fermi surface instability is as likely to occur as the PI. A long range interaction of Yukawa type, $V(r) = \frac{g}{r} e^{-\frac{r}{r_0}}$ on the other hand, does neither reveal any PI nor any topological Fermi surface instability.
\section{Conclusions and discussions}\label{conc}
In this paper, I have investigated the emergence and the nature of Pomeranchuk and topological transitions originating from a central electron-electron interaction on the surface of a 3D TI. The description is at the mean-field level and is in terms of a few microscopic parameters which are the angular momentum components of the interaction potential and their momentum space derivatives on the Fermi surface. This theory is applicable only to those 3D TIs for which the Fermi surface is nearly circular, and therefore such a continuum description remains valid.  The framework is very similar to that corresponding to the conventional SU(2) invariant Fermi liquids. However, qualitative as well as quantitative results are quite distinct owing to the presence of strong SOC in 3D TIs. The only similarity between the case of 3D TIs considered here and the conventional Fermi liquids is that the critical exponent in both the cases is 1/2, which is a characteristic of the mean-field approach. 
\paragraph*{}
 A microscopic expression of the PI equation  (\ref{insta}) corresponding to the interacting surface state of a 3D TI has been derived in terms of the parameters mentioned above. This equation has a few important implications which are markedly different from its SU(2) invariant counterpart occurring in conventional Fermi liquid. The PI equation signifies that the Fermi surface instability at a particular angular momentum channel $l$ can appear not only when the interaction strength in that channel exceeds the critical value $V_c$/2 but also when the interaction strength in any one of the $(l-1)$ and $(l+1)$ channels exceeds a value of $V_c$.  Therefore, although it is easier to achieve a Pomeranchuk phase transition in $l$'th channel when the interaction in the same is sufficiently strong, a combined effect of all the three angular momentum channels can drive the system into a symmetry broken phase without having interaction strength in any one of the channels larger than the critical value, as explained in FIG 3. However, the situation is indeed quite different when one considers the case of $l=2$ PI . In this case, even if the interaction strength in $l=1$ channel exceeds $V_c$/2, the instability is forbidden, and once the interaction exceeds $V_c$ in the same channel, the $l=2$ PI appears even if interaction in $l=2$ channel is absent. This is indeed quite remarkable and is a consequence of the absence of $l=1$ PI even in a SO coupled non-Galilean invariant Fermi liquid. For all the other angular momentum channels this is, in general, not true. For all other channels corresponding to $l\geq 3$ the Pomeranchuk order parameter corresponding to a particular channel, in general, depends on the competing interactions in both higher and lower angular momentum channels. Within the hierarchy assumption proposed here, any competing instabilities can be avoided up to the appearance of nematic instability in $l=2$ angular momentum channel since the $l=1$ PI is absent. The Pomeranchuk order parameter grows in a second order fashion with critical exponent 1/2. 
\paragraph*{}
Furthermore, it is shown that the phase separation instability corresponding to the $l=0$ channel can only appear for an attractive interaction of not only the density-density type but also a spin-charge or a pure spin-spin type. In particular, this is microscopically manifested by the finding that the interactions in any one or all of the angular momentum channels corresponding to $l\geq 0$ can drive the system towards instability. 
It is further shown that within the mean-field theory a hypothetical non-central interaction, which mimics the quadrupolar interaction, can produce nematic instability even with a sufficiently large but negative value of the interaction strength in $l=1$ angular momentum channel only. The ``\textit{hierarchy assumption}" introduced here has its effect in avoiding the only available competing PI coming from $l=3$ angular momentum channel, in absence of $l=1$ PI. The nematic order parameter emerging from the quadrupolar interaction is indeed proportional to the Pomeranchuk order parameter.
\paragraph*{}
Moreover, using a few realistic forms of e-e interaction it is shown that the PI can emerge from a repulsive interaction if and only if the interaction has a finite range $r_0$. However, the $l=2$ or the nematic instability can never occur when the interaction is repulsive. For repulsive interaction, on the other hand, instabilities in all the angular momentum channels with $l\geq 3$ can appear for finite range interactions. Surprisingly enough, the $l=4$ instability is more prone to appear than $l=3$ instability owing to the strong SOC. 
\paragraph*{}
In addition to the PI, the system exhibits another class of competing Fermi surface instability involving a change in the topology of the Fermi surface when the central e-e interaction is strictly repulsive. This is more of a reminiscent of the Lifshitz transition as explained in the section \ref{tpt}. It is further shown that there exist two types of topological FS instabilities at the most, viz., the appearance of vacated states at the centre, and a thin shell of vacated states within $k_F$. The topological phase transition reported here is driven by electron-electron interaction and turns out to be quite generic to the systems which support the emergence of the PI, and has also been obtained in the case of 3D Galilean invariant Fermi liquid \cite{QS}. Significantly enough, the topological FS instabilities appearing here for repulsive and central e-e interaction compete with the $l=3$, and $l=4$ PIs only. The nematic instability, on the other hand, is not competed by any of these topological FS instabilities.  The interaction induced Lifshitz/topological phase transition obtained in this paper from the theory of HFL is strictly valid only for strong TIs having a single disk type FS. However, recent findings of the appearance of such a topological phase transition in the doped Topological Crystalline Insulators strongly suggest that in strong TIs similar behaviours are expected to show up \cite{Ple, Gye}. On the physical ground, with the increase of electron number density (by doping), the electron-electron interaction strength increases which further is expected to lead to such interaction induced Lifshitz transition. Furthermore, this Lifshitz like transition must have some intriguing consequences in the surface transport properties which would be interesting to investigate in the future. The detection of entropy spikes can also serve as a signature of interaction driven Lifshitz transition \cite{Tsa}.
\paragraph*{}
 Moreover, signatures of the appearance of an annular FS, quite similar to what has been found here, has recently been discovered in a GaAs quantum well structure \cite{Wink}. Similar studies on the surface of 3D TIs are expected to reveal such topological FS instability from the topology of a disk to the topology of an annulus.
\section*{Acknowledgements}
Author acknowledges Yonatan Dubi for useful comments and suggestions. The author would like to express his appreciation for the valuale suggestions and criticisms of the anonymous referees in preparing the revised manuscript. 
\begin{widetext}
\begin{appendix}
\section{Detailed calculations of Landau Parameters}\label{app00}
One can calculate the expression for the projected Landau parameters (\ref{fl}) by multiplying both sides of (\ref{fl0}) by $(\frac{d \theta}{2\pi})$ and integrating from 0 to $2\pi$, and by using the parameterization (\ref{paramet}). Evaluating the integrals it is easy to find,
\begin{equation}
\bar{F}_{l} = \rho(\epsilon_{F}) \left[ \delta_{l,0} \bar{V} - \frac{1}{4}(1+\delta_{l,0}) V_{l} - \frac{1}{8} ( V_{1-l}+ V_{l-1} +V_{l+1}) \right].
\end{equation}
However, there are few facts corresponding to the above equation, viz., 
\begin{itemize}
\item[1.] when $l=1$, both $V_{1-l}$ and $ V_{l-1}$ are non-zero,\\ \item[2.] when $l=0$, only $V_{1-l} \neq 0$ but $V_{l-1} =0$, and \\ \item[3.] for $l\geq 2$, the quantity $V_{1-l}$ is always zero.
\end{itemize}
 These facts can be combined into a single expression where the term $(V_{1-l}+ V_{l-1})$ corresponding to the above equation can be replaced with $(1+\delta_{l,1}) V_{|l-1|}$ and equation (\ref{fl}) follows.
\section{Fermi velocity renormalization}\label{app0}
In absence of Galilean invariance one can determine the renormalization of the Fermi velocity by equating the total flux of the bare particles to that of the quasiparticles. The velocity operator for the bare particles is given by,
\begin{equation}\label{bare_vel}
v_{e} = v_{F}^{0}( \hat{z} \times \tau_{\sigma \sigma'}).
\end{equation}
By equating the total flux of the bare particles and and that of the quasiparticles, and projecting the Fermion operators on the Fermi surface corresponding to the Dirac cone with $\epsilon_{\mathbf{k}}>0$ (corresponding to the positive Helicity basis) one finds,
\begin{equation}\label{total_f}
\underbrace{\int \frac{d^2 k}{(2\pi)^2} (v_{F}^{0} \hat{\mathbf{k}}) \bar{n}_{\mathbf{k}}}_{\text{LHS=total flux of bare particles}} =\underbrace{ \int \frac{d^2 k}{(2\pi)^2} \bar{n}_{\mathbf{k}} [\nabla_{\mathbf{k}} \mathcal{E}_{\mathbf{k}}]}_{\text{RHS=total flux of quasi-particles}}.
\end{equation}
The next step is to evaluate the right hand side (RHS) of the above equation. The evaluation proceeds as follows, first one can notice that
\begin{eqnarray}\label{rhs0}
RHS &=& \int \frac{d^2 k}{(2\pi)^2} \bar{n}_{\mathbf{k}} \left[ v_{F} \hat{k} - \int \frac{d^2 k'}{(2\pi)^2} \nabla_{\mathbf{k}} \left( \frac{1}{2}(1+ \cos \theta_{k,k'}) \sum_{l=0}^{\infty} V_{l}(k,k') \cos (l\theta_{k,k'}) \bar{n}_{\mathbf{k'}} \right) \right],
\end{eqnarray} 
where the dispersion relation, $\mathcal{E}_{\mathbf{k}} = \epsilon_{\mathbf{k}} - \mu - \Sigma (\mathbf{k})$ has been used. Then the next three steps are, (i) to do an integration by parts of the second term of the above RHS, (ii) to use $\nabla_{\mathbf{k}} \bar{n}_{\mathbf{k}} = -\hat{k} \delta(|\mathbf{k}|-k_F)$, and (iii) to interchange $k$ and $k'$. Performing these steps one finds,
\begin{eqnarray}\label{rhs}
RHS &=& \int \frac{d^2 k}{(2\pi)^2} \bar{n}_{\mathbf{k}} \left[ v_{F} \hat{k} - \int \frac{d^2 k'}{(2\pi)^2} \hat{k'} \delta(|\mathbf{k}'|-k_F) \left( \frac{1}{2}(1+ \cos \theta_{k,k'}) \sum_{l=0}^{\infty} V_{l}(k,k') \cos (l\theta_{k,k'}) \right) \right].
\end{eqnarray}
Plugging (\ref{rhs}) into (\ref{total_f}) and scalar multiplying both sides of (\ref{total_f}) with $\hat{k}$ it can be easily found that,
\begin{eqnarray}
\int \frac{d^2 k}{(2\pi)^2} v_{F}^{0} \bar{n}_{\mathbf{k}}= \int \frac{d^2 k}{(2\pi)^2} v_{F} \bar{n}_{\mathbf{k}} &-& \int \frac{d^2 k}{(2\pi)^2} \bar{n}_{\mathbf{k}} \int \frac{k' d k' d\theta_{k'}}{(2\pi)^2} \cos (\theta_{k,k'}) \delta(|\mathbf{k}'|-k_F) \times \nonumber \\ & & \left( \frac{1}{2}(1+ \cos \theta_{k,k'}) \sum_{l=0}^{\infty} V_{l}(k,k') \cos (l\theta_{k,k'}) \right).
\end{eqnarray}
Evaluating the integral over $k'$ one finds,
\begin{equation}
\int \frac{d^2 k}{(2\pi)^2} v_{F}^{0} \bar{n}_{\mathbf{k}}= \int \frac{d^2 k}{(2\pi)^2}  \bar{n}_{\mathbf{k}} \left[ v_{F} - \frac{k_F}{16 \pi} \left( 2V_0 (k, k_F) +  2V_1 (k, k_F)+ V_2 (k, k_F) \right) \right],
\end{equation}
and equating the integrands corresponding to both the sides of the above equation on the Fermi surface ( i.e., for $k=k_F$) it turns out that,
\begin{equation}
v_{F}^{0} = v_{F} \left[ 1 - \frac{k_F}{16 \pi v_F}\left( 2V_0 (k_F, k_F) +  2V_1 (k_F, k_F)+ V_2 (k_F, k_F) \right) \right].
\end{equation}
On identifying $\bar{F}_1$ in the above equation, (\ref{vf}) corresponding to section \ref{mft} can be found. However, there is a subtlety in evaluating the integration by parts. In two dimensions $\nabla_{\mathbf{k}} = \hat{k} \frac{\partial}{\partial k} +\hat{\theta_k} \frac{\partial}{k \partial \theta_k} $, and it is easy to recognize that the $\theta_k$-integral coming from the $\frac{\partial}{ \partial \theta_k} $ term corresponding to (\ref{rhs0}) vanishes and only the terms corresponding to $\hat{k} \frac{\partial}{\partial k}$ remains. In this way evaluation of the integration by parts boils down to evaluate the following $k'$-integral,
\begin{equation}
\int k' dk' \bar{n}_{\mathbf{k'}} \, \frac{\partial [V_{l}(k,k')]}{\partial k'}  =2k_F V_{l}(k, k_F) - \int_{0}^{k_F} V_{l}(k,k') dk'.
\end{equation}
In the second term of the above equation, I now use the Taylor expansion of $V_{l}(k,k')$ around $k'=k_F$ and on the Fermi surface the only surviving term is $V_{l} (k,K_F)$. This leads to the value of the integration by parts to be $[k_F V_{l}(k,k_F)]$ which can be indeed be rewritten as, $[\int d^2 k' \delta(k'-k_F) V_{l}(k,k')]$ and has been used in (\ref{rhs}).
\section{Derivation of the self-consistency and instability equations}\label{app1}
In order to derive the self-consistency equation I use the following Taylor expansion:
\begin{equation}\label{Taylor}
V_{l}(k_F , k') = V_{l}(k_F , k_F) + \sum_{j=1}^{\infty} \frac{1}{j!} (k' - k_F)^{j} \; V_{l}^{(j)}(k_F, k_F), 
\end{equation}
in (\ref{selfcons}), where $V_{l}^{(j)}(k_F, k_F) = \left[\frac{\partial^{j} V_{l}(k_F, k')}{\partial k'^{j}}\right]_{k'=k_F} = V_{l}^{(j)}$, for all integer values of $j$, and for $j=0$ it is $V_{l}$. The integrals over $k'$ in the equation (\ref{selfcons}) then takes the form,
\begin{eqnarray}
\int_{k_F}^{k_F + \delta k_{F}(\theta_{k'})} k' dk' V_{l}(k_F , k') &=& V_{l}(k_F , k_F) \left(k_F \delta k_{F}(\theta_{k'}) + \frac{[\delta k_{F}(\theta_{k'})]^2}{2} \right) \nonumber \\ &+& \sum_{j=1}^{\infty} \frac{1}{j!} V_{l}^{(j)}(k_F, k_F) \left[ k_F \frac{(\delta k_{F}(\theta_{k'}))^{j+1}}{j+1} + \frac{(\delta k_{F}(\theta_{k'}))^{j+2}}{j+2} \right],
\end{eqnarray}
when the above Taylor expansion is used. Further using the above equation in (\ref{selfcons}) one can easily calculate the following,
\begin{eqnarray}
\Sigma_{l} = \int_{0}^{2\pi} \frac{d \theta_{k'}}{4\pi^2} \cos(l \theta_{k'})\Bigg[ \frac{1}{4} \mathcal{V}_{l} \left(k_F \delta k_{F}(\theta_{k'}) + \frac{[\delta k_{F}(\theta_{k'})]^2}{2} \right) + \sum_{j=1}^{\infty} \frac{1}{j!} \mathcal{V}_{l}^{(j)} \left[ k_F \frac{[\delta k_{F}(\theta_{k'})]^{j+1}}{j+1} + \frac{[\delta k_{F}(\theta_{k'})]^{j+2}}{j+2} \right],
\end{eqnarray}
where the self-consistency of the above equation comes from the equation (\ref{delkf}), and $\mathcal{V}_{l}$ has already been defined as $\mathcal{V}_{l} = [2V_l + (1+\delta_{l,1} ) V_{|l-1|} + V_{l+1}]$. To lowest order in $\delta k_{F}(\theta_{k'})$, i.e., for $j=1$ in the Taylor expansion (\ref{Taylor}) one finds, after doing the integration over $d \theta_{k'}$ corresponding to the above equation,
\begin{equation}
\mathcal{V}_{l} = \frac{16 \pi v_F}{k_{F}},
\end{equation}
which is nothing but equation (\ref{insta}), i.e. the PI condition. Similarly, to next higher order, i.e., $j=2$, and next to next higher order, i.e., $j=3$, in the Taylor expansion mentioned above, one finds (\ref{nh1}) and (\ref{insta_1}) respectively.
\end{appendix}
\end{widetext}

\end{document}